\documentclass[superscriptaddress,showpacs,amssymb,10pt,reprint,aps,prd,longbibliography,footinbib,floatfix]{revtex4-1}

\usepackage{graphicx,epsfig,amssymb} 
\usepackage{amsmath,amsfonts, times}
\usepackage{bm} 
\usepackage{epstopdf}
\usepackage[linktocpage,colorlinks]{hyperref}
\usepackage[caption=false]{subfig}
\usepackage[usenames]{color}     
\usepackage{natbib}
\usepackage{subfig}
\usepackage[utf8x]{inputenc}

\usepackage{soul}
\usepackage{xcolor}

\begin{document}
\title{\large Spectral lines of dirty wormholes}

\author{Leonardo K. S. Furuta}
\email{leonardo.furuta@icen.ufpa.br}
\affiliation{Programa de P\'os-Gradua\c{c}\~{a}o em F\'{\i}sica, Universidade 
	Federal do Par\'a, 66075-110, Bel\'em, Par\'a, Brazil.}

\author{Renan B. Magalh\~aes}
\email{magalhaes.renan.b@gmail.com}
\affiliation{Departamento de F{\'i}sica, Universidade Federal do Maranh{\~a}o, 65080-805, S{\~a}o Lu{\'i}s, Maranh{\~a}o, Brazil.}

\author{Haroldo C. D. Lima Junior}
\email{haroldo.lima@ufma.br}
	\affiliation{Departamento de F{\'i}sica, Universidade Federal do Maranh{\~a}o, 65080-805, S{\~a}o Lu{\'i}s, Maranh{\~a}o, Brazil.}

\author{Lu\'is C. B. Crispino}
\email{crispino@ufpa.br}
\affiliation{Programa de P\'os-Gradua\c{c}\~{a}o em F\'{\i}sica, Universidade 
	Federal do Par\'a, 66075-110, Bel\'em, Par\'a, Brazil.}

\begin{abstract}
Astrophysical objects like black holes are usually surrounded by matter in the form of accretion disks or jets of matter. These astrophysical scenarios are expected to introduce novel phenomenology in the scattering of particles and fields. Wormholes are viable candidates for exotic compact objects that can mimic some black hole properties. Hence, it is natural to wonder what would happen if the central astrophysical object were a wormhole, instead of a black hole. We investigate the astrophysical environment effect on the absorption of a massless scalar field by a dirty wormhole surrounded by a thick shell of matter. We study null geodesics around these dirty wormholes and analyze under which conditions new pairs of light rings can appear. The presence of new stable light rings allows new quasibound states in the spacetime, apart from the ones trapped near the throat. Thus, the astrophysical environment can introduce deviations in the absorption bands.
Remarkably, although heavy and dense distributions of matter are considered surrounding the wormhole, the position of most of the spectral lines in the absorption bands is preserved, indicating that the astrophysical environment cannot hide some fingerprints of the central object.
\end{abstract}

\maketitle

\section{Introduction}\label{sec:int}
Different types of compact objects are allowed by General Relativity (GR). Among them, there are black holes, which have played theoretical and observational major role in understanding gravity. In particular, they have been used to probe strong gravity effects in astrophysical scenarios, and are regarded as the main candidates to explain the experimental data coming from gravitational wave detections~\cite{Abbott1:2016, Abbott2:2016, Abbott1:2017, Abbott2:2017} and the shadow images from very long baseline interferometry (VLBI)~\cite{EHT:2019}. 

Within GR, the most important family of black hole solutions, the Kerr-Newman black holes, arises from the coupling of Einstein's gravity to Maxwell's electromagnetism. Solutions within this family are entirely described by their mass, charge and angular momentum, and, from the Kerr hypothesis, provide our current best description of black holes. However, astrophysical black holes are not believed to be completely isolated in the Universe, such that matter distributions in the form of disks or jets of matter are expected. Therefore one has to consider the interaction of black holes with their environment. 

In this regard, the so-called dirty black holes were proposed~\cite{Visser:1992}. \textit{Dirtiness} stands for fields or matter distributions interacting with black holes. Dirty black holes arise, for example, from the coupling between gravity and electromagnetism to additional fields such as axionic~\cite{Bardoux:2012}, dilatonic~\cite{Gibbons:1988}, Abelian~\cite{Achucarro:1995} and non-Abelian fields~\cite{Volkov:1999}. An instructive case of dirty black hole, which has gained some attention in the literature, is constructed via the introduction of either thin or thick shells of matter surrounding the central black hole. The advantage of these solutions is the possibility to model configurations of ordinary, phantom and even dark matter surrounding the central object~\cite{Konoplya:2019}.

The presence of matter around compact objects may introduce new phenomenology, for example deviations in their shadow and gravitational lensing or in their quasinormal mode spectrum. This suggests that the investigation of the effect of a non-empty environment near objects beyond black holes is relevant, since the environment could spoil some of the \textit{fingerprints} that tell these objects apart from black holes. Some alternatives beyond the black hole paradigm are exotic compact objects (ECOs) without horizon, such as boson stars~\cite{kaup1968klein}, wormholes~\cite{Salcedo:2021}, fuzzballs~\cite{Skenderis:2008}, and many others~\cite{Cardoso:2019}. Even though many of these ECOs can mimic black hole characteristics, they are known for presenting distinct phenomenology, what could introduce traits that allow to differentiate them from black holes. These characteristics could be affected by their environment, what motivates a study of environmental effects on ECOs. 

Wormholes, amid the myriad of ECOs present in the literature, have been receiving increasing attention in theoretical physics. These objects with nontrivial topology, often identified by a hypersurface of minimal area, may enable shortcuts between separated places or times, either in the same or different universes. Regrettably, in (4 dimensional) GR, traversability usually does not come without the violation of some energy conditions~\cite{Morris2:1988}, and usually exotic matter is required to support stable and traversable wormholes. In this context, the thin-shell formalism is useful for constraining violations of certain energy conditions, exclusively to the wormhole throat. This formalism consists in the grafting of two spacetimes at a given hypersurface. One of the simplest thin-shell wormholes is obtained by matching two copies of the Schwarzschild spacetime at a hypersurface outside the Schwarzschild radius $r=2M$. Depending on the compactness of the wormhole, the throat may be located inside the photon spheres inherited from the Schwarzschild spacetimes. When these ultracompact wormholes are perturbed by waves, quasibound states are expected to arise, which are known for introducing characteristic signatures in the scattering analysis. These quasibound modes may lead to a series of echoes in their ringdown profile at late times and to resonant peaks in their absorption spectra. These distinct phenomenologies could, in principle, be measured in future spectroscopy experiments, and therefore the investigation of how the environment can affect these features should be addressed.

Environmental effects were considered in this particular thin-shell wormholes spacetime in Ref.~\cite{KZ:2019}. In that work the authors focused on how the surrounding matter affects the echoes of scalar perturbations. Here, we investigate how the surrounding matter, in the form of a thick shell, influences the absorption of massless particles and scalar waves in dirty wormholes. The model we are considering consists of a dirty Schwarzschild spacetime surrounded by a thick shell of matter connected to a standard Schwarzschild spacetime. In particular, we study some null geodesics around these dirty wormholes and analyze in which conditions new light rings can appear. Remarkably, when new light rings appear they do it in pairs, one of them being stable~\cite{xavier2024traversable}. The presence of novel stable photon spheres may enable new trapped modes on the spacetime, what reflects on deviations in the absorption bands. We analyze such deviations in detail and discuss the persistence of wormhole fingerprints even when surrounded by heavy and dense distributions of matter.

The paper is organized as follows. In Sec.~\ref{sec:tswh} we briefly review the thin-shell formalism in GR and perform a grafting of a dirty Schwarzschild spacetime surrounded by a thick shell of matter and a Schwarzschild spacetime in empty space. In Sec.~\ref{sec: null} we study the null geodesic structure of these dirty wormholes. In Sec.~\ref{sec:abs} we investigate massless scalar waves in the vicinity of dirty wormholes, discussing quasibound states and how they affect the absorption spectra. Our final remarks are presented in Sec.~\ref{sec: FR}. We use natural units, such that $G = \hbar = c =1$, and adopt the metric signature $(-,+,+,+)$.

\section{Dirty wormholes}\label{sec:tswh}
We are interested in investigating how the astrophysical environment can affect the absorption properties of wormholes in GR. The model considered here, and proposed in Ref.~\cite{KZ:2019}, consists of a thin-shell Schwarzschild wormhole with a thick layer of matter covering the throat, henceforth dubbed dirtiness. Before we specify the shape of the dirtiness-mass function, let us review the construction of thin-shell wormholes in GR. 

\subsection{Thin-shell wormholes in GR}
Traversable wormholes constructed within GR (in 4 dimensions) usually violate some energy conditions, such that one requires distributions of exotic matter to support stable throats~\citep{Morris1:1988}. A possibility to diminish the amount of exotic matter is to build wormholes via the thin-shell formalism~\cite{Visser:1989}. This formalism consists in the grafting of two manifolds (${\cal M}_+$ and ${\cal M}_-$) at a thin hypersurface $\Sigma$, called shell. In order to make this gluing technique mathematically well-posed one has to work with tensorial distributions instead of standard tensorial functions. With these ingredients, one can compute the allowed discontinuities in the geometrical quantities. This is translated in the so-called junction conditions. 

In the thin-shell formalism, the geometric and matter tensorial functions are replaced by tensorial distributions. In this regard, the information of the metric tensor in each manifold ($g_{\mu\nu}^{+}$ in ${\cal M}_+$ and $g_{\mu\nu}^{-}$ in ${\cal M}_-$) is then put in a single tensorial distribution~\cite{Clarke:1987}
\begin{equation}
\underline{g}_{\mu\nu} = g_{\mu\nu}^{+}\underline{\theta} + g_{\mu\nu}^{-}(\underline{1} - \underline{\theta}),
\end{equation}
where the underline is used to denote distributions and $\underline{\theta}$ is the Heaviside step function, which takes the value 0 in ${\cal M}_-$, 1 in ${\cal M}_+$ and a reference value (for instance, 1/2) on the hypersurface $\Sigma$. Analogously, the distributional form of the determinant of the metric can be regarded as $\underline{|g|} = |g|^{+}\underline{\theta} + |g|^{-}(\underline{1} - \underline{\theta})$, where $|g|^+$ and $|g|^-$ stand, respectively, for the determinant of $g_{\mu\nu}^{+}$ and $g_{\mu\nu}^{-}$.
The metric, like its determinant, is assumed to be continuous through the shell, what is achieved by requiring $g_{\mu\nu}^{+}\vert_{\Sigma}=g_{\mu\nu}^{-}\vert_{\Sigma}$. By introducing the bracket notation for the discontinuities, namely $[A] \equiv A^+\vert_{\Sigma} - A^-\vert_{\Sigma}$, one can express the continuity of the metric and its determinant as $[g_{\mu\nu}]=0$ and $[|g|]=0$, respectively.

In this approach, the matter content is contained in the energy-momentum tensor distribution
\begin{equation}
\underline{T}_{\mu\nu} = T_{\mu\nu}^{+}\underline{\theta} + T_{\mu\nu}^{-}(\underline{1} - \underline{\theta}) + S_{\mu\nu}\underline{\delta}^\Sigma,
\end{equation}
where $S_{\mu\nu}$ is the singular part of the energy-momentum (the matter content on the hypersurface) and $\underline{\delta}^{\Sigma}$ is a Dirac's delta-type distribution with support on the hypersurface defined through $<\underline{\delta}^\Sigma,X> \equiv \int_{\Sigma} X$, for any ${\cal C}^1$ function $X$ of compact support~\cite{Mars:1993}. In the same manner, the trace of the energy momentum tensor reads $\underline{T} = T^{+}\underline{\theta} + T^{-}(\underline{1} - \underline{\theta}) + S\underline{\delta}^\Sigma$, where $T^+$ and $T^-$ denote, respectively, the trace of the energy-momentum tensor in ${\cal M}_+$ and ${\cal M}_-$, and $S\equiv S^\alpha_{\ \ \alpha}$ the trace of the singular part of the energy-momentum tensor.

Even though the metric is continuous across the shell, other matter and geometrical quantities do not have to be. For instance, the energy-momentum tensor and its trace, if one combines two asymmetric matter sectors~\cite{magalhaes2023asymmetric}. To determine the allowed discontinuities of those quantities one has to use the junction conditions, namely a set of equations that relate the discontinuities of curvature and matter quantities. The junction conditions are highly dependent on the gravitational framework one is considering. In GR these conditions were established by Darmois and Israel~\cite{darmois1927memorial,Israel:1965,Senovilla:2013}, namely 
\begin{align}
\label{eq:h_cont}&[h_{\mu\nu}]=0,\\
\label{eq:Ktrace}&h_{\mu\nu}[K^{\alpha}_{\ \alpha}]-[K_{\mu\nu}] = \kappa^2 S_{\mu\nu},\\
&(K^+_{\mu\nu}+K^-_{\mu\nu})S^{\mu\nu}=2n^\mu n^\nu [T_{\mu\nu}],\\
\label{eq:covD_shell}&D^\mu S_{\mu\nu}= -n^\mu h^{\alpha}_{\nu}[T_{\alpha\mu}].
\end{align}
Here $h_{\mu\nu} = g_{\mu\nu} - n_{\mu}n_\nu$ and $K^{\pm}_{\mu\nu} = h^{\alpha}_{\ \mu} h^{\beta}_{\ \nu}\nabla^{\pm}_{\alpha}n_\beta$ denote, respectively, the first fundamental form (the induced metric on the hypersurface) and the second fundamental form (the extrinsic curvature of the shell), and $n_{\mu}$ is the normal vector to the hypersurface $\Sigma$. In Eq.~\eqref{eq:covD_shell}, $D^\mu \equiv h^{\mu}_{\ \nu} \nabla^\nu$ is the intrinsic covariant derivative within the shell. Equation~\eqref{eq:Ktrace} implies that the brane tension in general is nonvanishing, $\kappa^2 S = 2[K]$. 
As previously mentioned, the junction conditions are highly dependent on the gravity theory we are dealing with. In modified gravity scenarios, the above set is in general replaced by another one. This allows, for example, one to build stable thin-shell wormholes supported by positive energy densities, then satisfying energy conditions in the whole spacetime~\cite{Merce:2021}.

\subsection{Thin-shell Schwarzschild wormholes}
With the junction conditions~\eqref{eq:h_cont}-\eqref{eq:covD_shell}, one can graft two Schwarzschild spacetimes above their horizons, in order to construct a traversable wormhole. Hence, we consider two Schwarzschild spacetimes, whose line elements are given by
\begin{equation}
\label{eq:schw_+-} ds^2_{\pm} = -f_{\pm}(r_{\pm})dt^2+f_{\pm}^{-1}(r_{\pm})dr_{\pm}^2+r_{\pm}^2d\Omega^2,
\end{equation}
where $r_\pm$ denotes, respectively, the radial coordinate in ${\cal M}_+$ and ${\cal M}_-$, $d\Omega^2\equiv d\theta^2 + \sin^2\theta d\phi^2$ is the line element of a unit sphere and $f_{\pm}(r_\pm) = 1-2M/r_\pm$, in the absence of dirtiness. We match the two Schwarzschild spacetimes 
at a shell of radius $r_\pm=r_0$. 
The location of the horizon is given by the roots of $f(r_\pm)$, that is $r_\pm = 2M$. Therefore, the shell location (the throat radius) must be positioned outside the event horizon in order to guarantee that the graft procedure results in a traversable wormhole, i.e. $r_0>2M$.

The energy-momentum tensor distribution of the resulting spacetime reads
\begin{equation}
\underline{T}^{\mu\nu} = S^{\mu\nu}\underline{\delta}^{\Sigma},
\end{equation}
where, in the forthcoming analysis, we will consider that the matter content on the shell is modeled by a perfect fluid, $S^{\mu\nu}= \text{diag}(-\sigma,p,p)$, where $\sigma$ is the surface energy density and $p$ is the tangential surface pressure. It can be shown that $\sigma$ and $p$ of the thin shell of matter are given, respectively, by~\cite{Cardoso:2016}
\begin{equation}
\label{eq: SP} \sigma = -\dfrac{1}{2\pi r_{0}}\sqrt{f(r_0)}, \ \ \ \ \ p = \frac{1}{4\pi r_{0}}\frac{f(2r_0)}{\sqrt{f(r_0)}}.
\end{equation} 


\subsection{Dirty wormholes}
Astrophysical objects are generally observed to be surrounded by matter distributions, for instance the hot plasma around supermassive black holes~\cite{Tursunov:2020}. One expects that the presence of matter near compact objects introduces some new phenomenology that can be used to tell them apart from the idealized scenarios without matter in the environment. A first approach is to introduce a thin shell of matter around compact objects, like, for example, the dirty black holes studied in Ref.~\cite{Caio:2016}. However, these dirty black holes present discontinuities in the metric tensor.  Here, on the other hand, we consider a thick shell of matter outside our target object, giving rise to a dirty wormhole. To do so, in the outer universe of the thin-shell Schwarzschild wormhole, we introduce a mass function $m(r_{+})$, such that
\begin{equation}
\label{thick_shell}\quad f_+(r_+)\rightarrow 1-\frac{2 m(r_+)}{r_+}.
\end{equation}
The explicit form of the mass function $m(r_{+})$ depends on the model of the thick shell of matter we are considering. Here, we follow the procedure of Ref.~\cite{KZ:2019}, and chose the mass function as
\begin{equation}
\label{mass_profile} m(r_+) = \begin{cases} M, &r_+ < r_{s},\\
M + R(r_+,r_{s},\Delta r_{s}) \Delta M,  &r_{s} \leq r_+ \leq r_{s} + \Delta r_{s},\\
M + \Delta M,  &r_+ > r_{s} + \Delta r_{s},
\end{cases}
\end{equation}
with
\begin{equation}
R(r_+,r_{s},\Delta r_{s}) = \left(3 - 2\frac{r_+ - r_{s}}{\Delta r_{s}} \right) \left( \frac{r_+ - r_{s}}{\Delta r_{s}}\right)^{2},
\end{equation}
where $r_{s}$, $\Delta r_{s}$ and $\Delta M$ are the radius, thickness and mass of the shell, respectively. As pointed out in Ref.~\cite{KZ:2019}, the metric function \eqref{thick_shell} is of class ${\cal C}^1$ for a thick shell described by the mass term~\eqref{mass_profile}.
Throughout this paper the thick shell of matter is assumed to be outside the innermost stable circular orbit (ISCO) of the Schwarzschild spacetime, i.e., $r_{s} \geq 6M$. In order to prevent the existence of horizons outside thick shell, we also demand that the upper limit of the thick shell, $r_s+\Delta r_s$, lies outside the Schwarzschild radius of a configuration with net mass $M+\Delta M$, that is $r_s+\Delta r_s>2(M+\Delta M)$.

We consider matter surrounding the thin-shell wormhole with either positive or negative energy density. Even though the phantom matter (negative energy density distribution) is less astrophysically relevant than the normal matter (positive energy density distribution), it is sometimes used as the source of wormhole geometries~\cite{Sushkov:2005}. 

\subsection{Embedding diagrams}
In order to visualize the wormhole structure, one can perform the embedding, in the equatorial plane and at a constant $t$, of the wormhole in a 3D Euclidean space. 

First, we note that the radial coordinates $r_{\pm}$ cannot be used as global coordinates, since 
$r_+$ ($r_-$) describes solely the $\mathcal{M}_{+}$ ($\mathcal{M}_{-}$) side of the throat.
 It is useful, here and in the forthcoming analysis, to define a global coordinate $r_{\star}\in (-\infty, \infty)$, implicitly given by	
\begin{equation}
\label{eq: Tortoise} dr_{\star} = \pm\dfrac{dr_\pm}{f_\pm(r_\pm)},
\end{equation} 
where the positive (negative) sign describes the upper (lower) side of the wormhole throat. We choose an integration constant such that $r_{\star}=0$ corresponds to the wormhole throat.

\begin{figure*}
\includegraphics[width=\columnwidth]{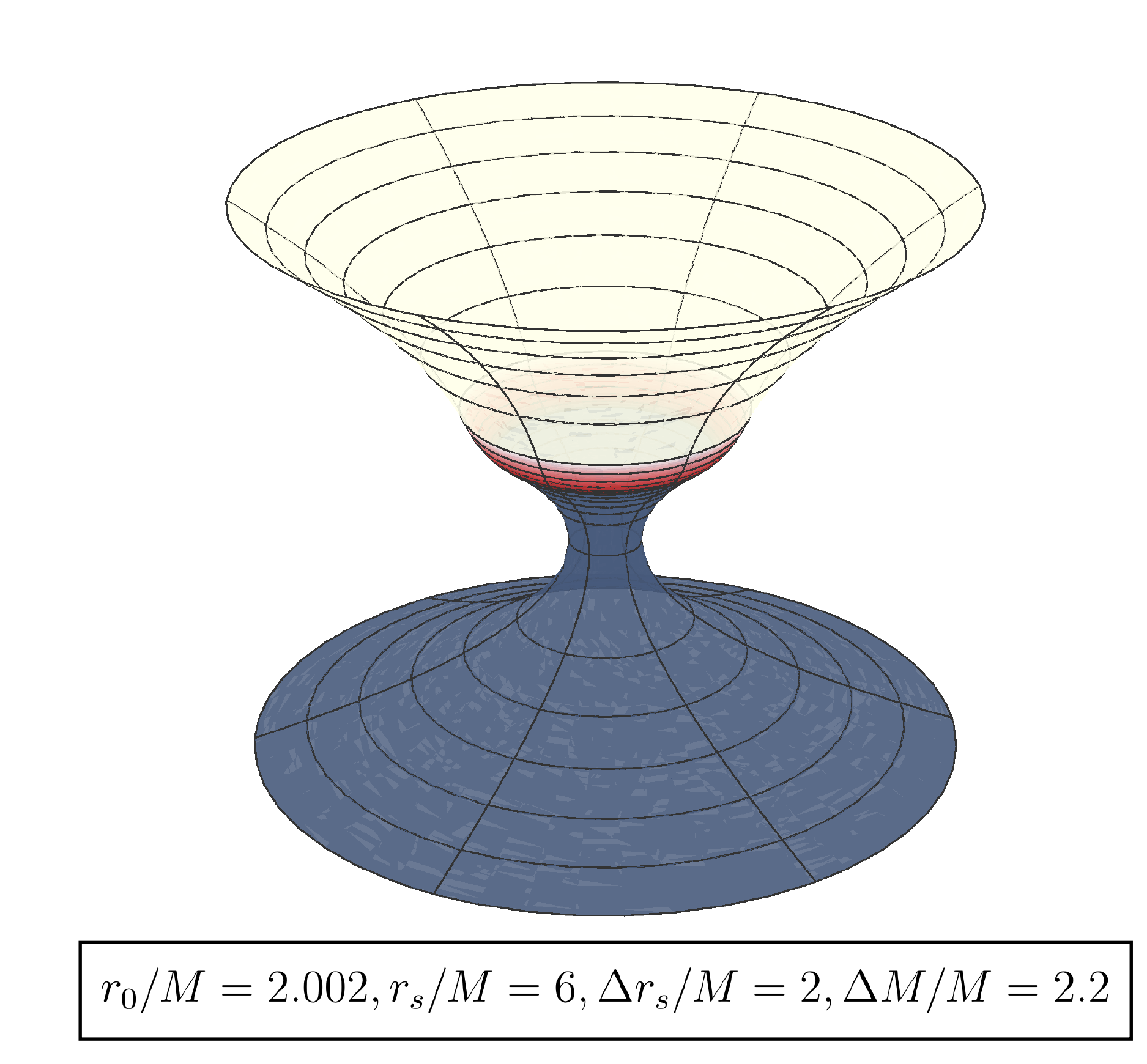}\includegraphics[width=\columnwidth]{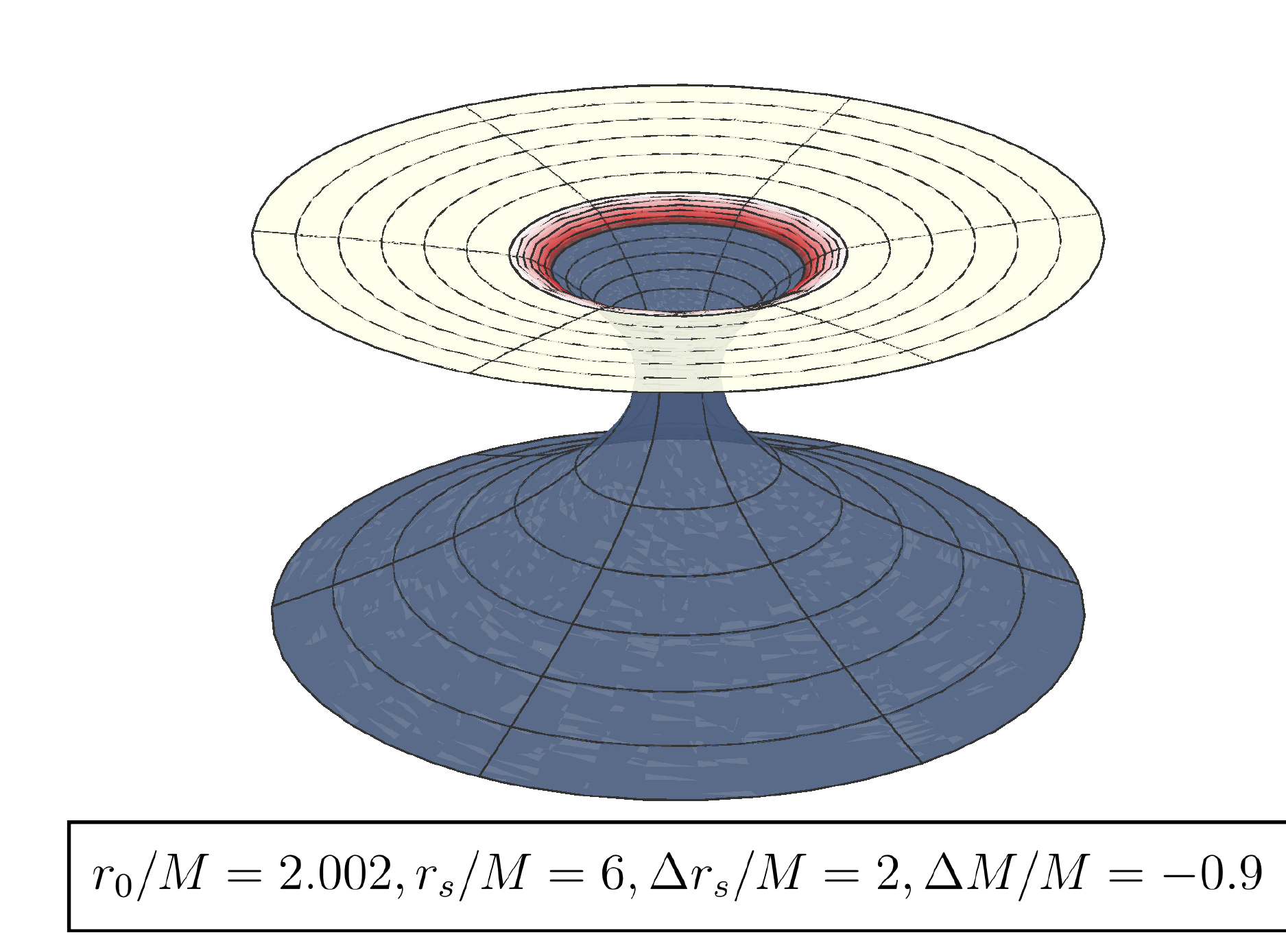}
\caption{Embedding diagrams of dirty wormholes. The left panel depicts a dirty wormhole surrounded by normal matter, while the right panel represents a dirty wormhole surrounded by phantom matter.}
\label{fig:embeddings}
\end{figure*}

In terms of the global coordinate $r_{\star}$, the line element of the 2D surface to be embedded is
\begin{equation}
\label{eq: ScwhEquatorial1} ds^2 = \begin{cases} f_{+}(r_{\star})dr_{\star}^2 + r_+^{2}(r_{\star})d\phi^{2}, &r_\star\geq0,\\
f_{-}(r_{\star})dr_{\star}^2 + r_-^{2}(r_{\star})d\phi^{2}, &r_\star<0.
\end{cases}
\end{equation}
The embedding is then performed in a 3D Euclidean space, that in cylindrical coordinates is
\begin{equation}
\label{eq: EuclidCyli1} ds^{2} = d\rho^{2} + \rho^{2}d\phi^{2} + dz^{2}.
\end{equation}
By allowing that $\rho$ and $z$ to be functions of $r_{\star}$ and comparing Eq.~\eqref{eq: ScwhEquatorial1} with Eq.~\eqref{eq: EuclidCyli1}, one obtains, in each side of the throat,
\begin{equation}
\label{eq: ODEZ1} \left(\dfrac{dz}{dr_\star}\right)^2  = \begin{cases}
f^2_{+}(r_\star)\left(\dfrac{1}{f_{+}(r_\star)}-1\right), &r_\star\geq 0,\\
f^2_{-}(r_\star)\left(\dfrac{1}{f_{-}(r_\star)}-1\right), &r_\star< 0, \end{cases}
\end{equation}
where we have taken $\rho = r_+$, if $r_\star\geq 0$ and $\rho = r_-$, if $r_\star< 0$. It is important to point out that we can only embed the 2D surface~\eqref{eq: ScwhEquatorial1} in the 3D Euclidean space whether $1/f_{\pm}>1$, otherwise $dz/dr_\star$ is either imaginary or identically vanishes. Throughout this paper we are considering the condition $z(0) = 0$. In Fig.~\ref{fig:embeddings} we exhibit the embedding diagrams of two dirty wormhole configurations. We point out that configurations surrounded by phantom matter cannot be entirely embedded in 3D Euclidean space when $\Delta M<-M$, since $1/f_+<1$ outside the thick shell in this case.

\section{Null geodesics}\label{sec: null}
The dirty wormhole considered in this paper is a spherically symmetric geometry, thus we investigate, without loss of generality, the null geodesic flow at the equatorial plane $(\theta=\pi/2)$. The null geodesic analysis is important since it is the high-frequency limit of our full numerical results~\cite{Haroldo:2020}, presented in Sec.~\ref{sec:abs}. Moreover, the null geodesic analysis is related to the properties of light rings, shadows and gravitational lensing in such wormhole geometries~\cite{Abdujabbarov:2016,Ohgami:2015,Nandi:2006}.

The Lagrangian $\mathcal{L}_{\text{geo}}$, describing the null geodesics at the equatorial plane, in each side of the throat, reads
\begin{align}
\label{lagrangian}\,\mathcal{L}_{\text{geo}}=\begin{cases} -f_{+}(r_\star)\, \dot{t}^2+f_+(r_\star)\dot{r}_\star^2+r_+^2(r_\star)\dot{\phi}^2, &r_\star\geq 0,\\
-f_-(r_\star)\, \dot{t}^2+f_-(r_\star)\dot{r}_\star^2+r_-^2(r_\star)\dot{\phi}^2, & r_\star <0,
\end{cases}
\end{align}
where $\mathcal{L}_{\text{geo}}=0$, once that it describes null geodesics. Since the metric is continuous across the throat, it follows that the Lagrangian is also continuous. 
The inspection of Eq.~\eqref{lagrangian}, reveals that $t$ and $\phi$ are cyclic coordinates from the standard point of view of classical mechanics. Thus, the conjugated momenta $p_t=-E$ and $p_\phi=L$, respectively, are conserved along geodesic motion. Moreover, the Euler-Lagrange equation for the cyclic coordinates results in first order equations of motion:
\begin{align}
\label{tdot}&\dot{t}=\begin{cases}
E/f_{+}(r_\star), &r_\star\geq 0,\\
E/f_{-}(r_\star), &r_\star< 0,\\
\end{cases} \\
\label{phidot}&\dot{\phi}=\begin{cases}
L/r_+^2(r_\star), &r_\star\geq 0,\\
L/r_-^2(r_\star), &r_\star< 0.\\
\end{cases}
\end{align}
By substituting Eqs.~\eqref{tdot}-\eqref{phidot} in Eq.~\eqref{lagrangian}, we obtain that null geodesics satisfy, in each side of the throat,
\begin{align}
\label{xdot}&f_\pm^2(r_\star)\dot{r}_\star^2+V_{\text{geo}}(r_\star)=E^2,
\end{align}
where the effective potential for null geodesics in the dirty wormhole spacetime is given by
\begin{align}
\label{veff_geo} V_{\text{geo}}(r_\star)=\begin{cases}
\dfrac{f_+(r_\star)\,L^2}{r^2_+(r_\star)}, &r_\star\geq 0,\\
\dfrac{f_-(r_\star)\,L^2}{r^2_-(r_\star)}, &r_\star< 0.\\
\end{cases}
\end{align}

\begin{figure*}[h!]
\includegraphics[width=\columnwidth]{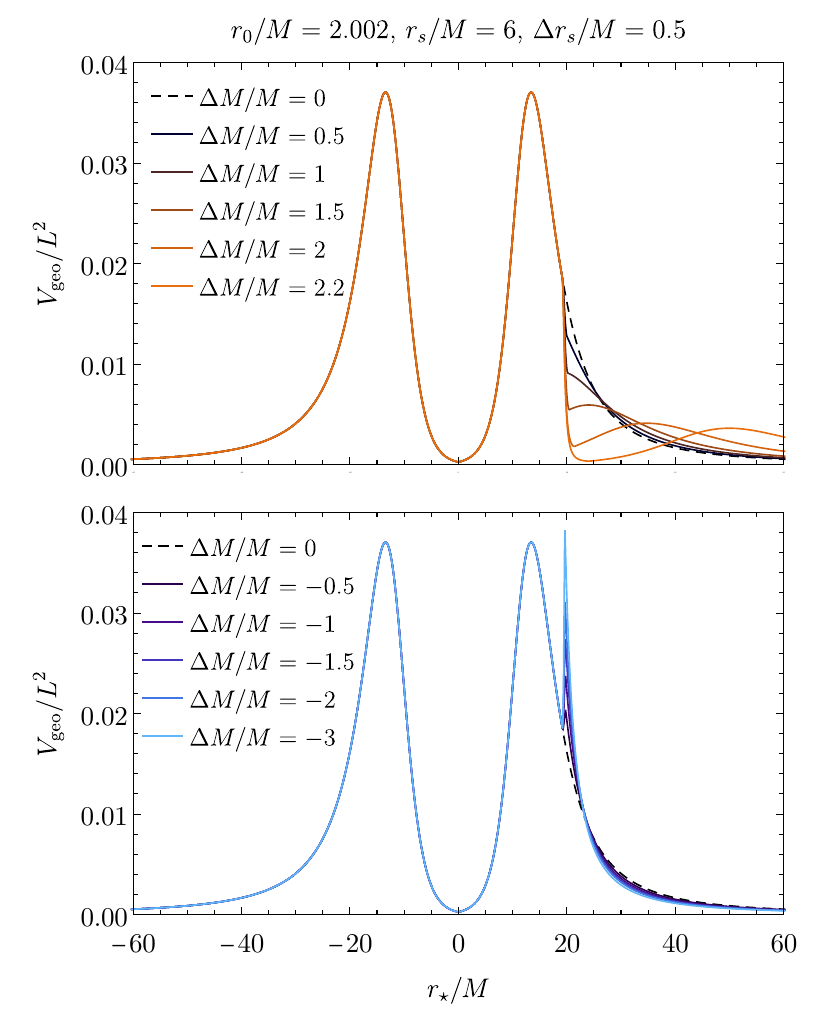}\includegraphics[width=\columnwidth]{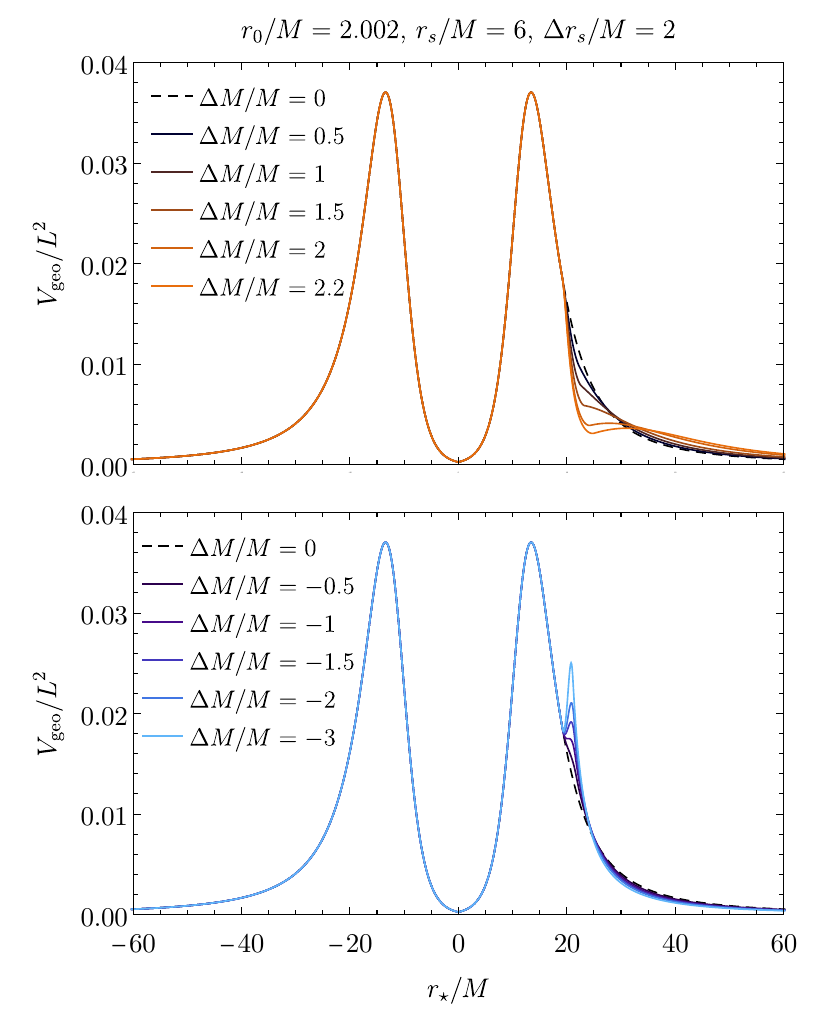}\\\includegraphics[width=\columnwidth]{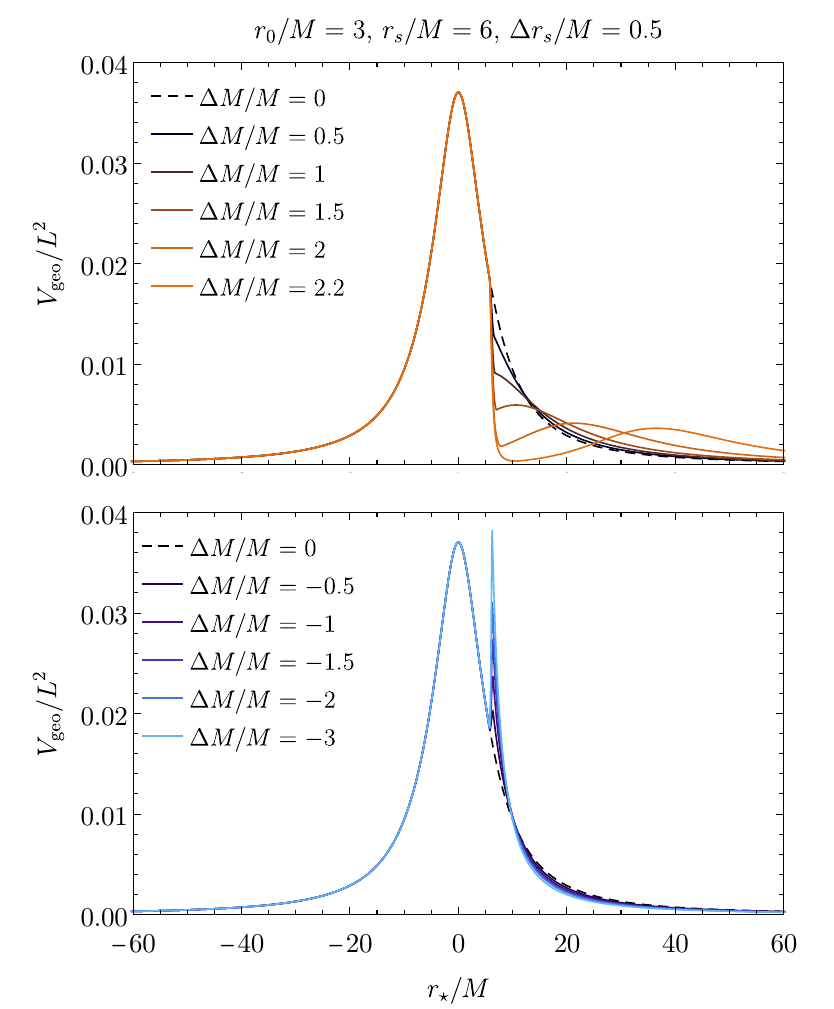}\includegraphics[width=\columnwidth]{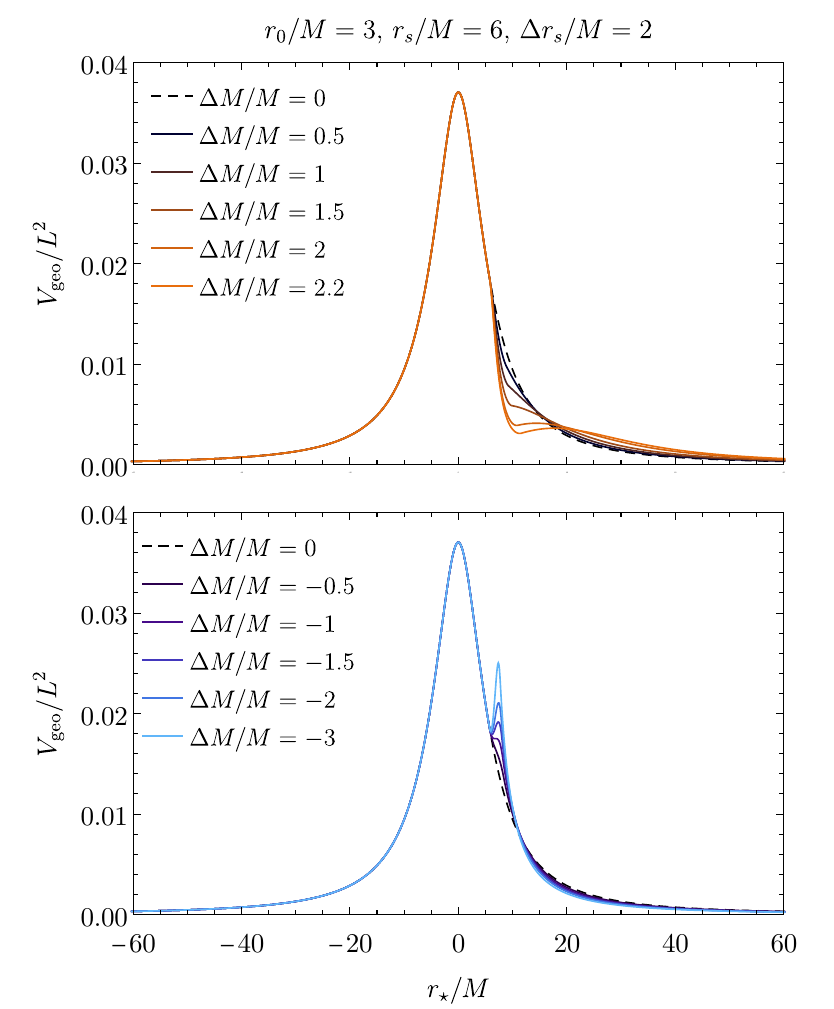}
\caption{Effective potential for null geodesics of some dirty wormhole configurations. The top panels of each quadrant (plots with orange-like color scheme) correspond to thick shells with positive energy density, while the bottom panels of each quadrant (plots with blue-like color scheme) depict thick shell configurations with negative energy density. For comparison we also plot the corresponding thin-shell Schwarzschild wormhole case (black dashed line).}
\label{VgeoM}
\end{figure*}

A careful analysis of the effective potential behavior enables us to extract information about the structure of the dirty wormhole. In Fig.~\ref{VgeoM}, we exhibit the effective potential as a function of the global coordinate $r_\star$ for some dirty wormhole configurations and compare with the corresponding thin-shell Schwarzschild wormhole without surrounding matter. 
In the configurations shown in the top rows of each quadrant of Fig.~\ref{VgeoM}, the throat is located inside the photon sphere of the Schwarzschild spacetime and outside the sphere with radius $2M$, namely $2M<r_0=2.002M<3M$. Hence, two unstable light rings (local maxima in the effective potential), inherited from the Schwarzschild spacetime, are present, one in each side of the throat. Between these unstable light rings, there is also a stable light ring (local minima in the effective potential) located at the throat. On the other hand, in the bottom rows of each quadrants of Fig.~\ref{VgeoM}, the gluing of the two Schwarzschild spacetimes is made precisely at their photon spheres. In this case, the effective potential presents two unstable light rings and one stable light ring.

However, as one can see from Fig.~\ref{VgeoM}, these light rings due to the gluing of two Schwarzschild spacetimes are not the only extreme points in the effective potential, and the presence of the thick shell of matter also introduce new pairs of light rings in the spacetime. One can see that the presence of shells with positive energy density introduces a well in the effective potential that is larger and deeper as the amount of matter increases. Just after the valley, a new local maximum arises before the effective potential vanishes asymptotically. Conversely, the presence of matter with negative energy density introduces an additional narrow valley followed by a narrow peak in the effective potential. The width of the new peak is almost insensitive to the change in the amount of matter in the shell, and, as one increases $|\Delta M|$, the height increases. Moreover, we notice that for more dilute shells, i.e. configurations with larger $\Delta r_s$, the new peaks and valleys are less prominent and the effect of the shell of matter in the effective potential is more attenuated. For instance, by fixing $\Delta M>0$, as one increases $\Delta r_s$, the valleys become shallower. Reciprocally, by fixing $\Delta M<0$, as one increases $\Delta r_s$, the height of the additional peak diminishes. If one spreads the shell in the outer universe, $\Delta r_s\to \infty$, $R(r_+,r_{s},\Delta r_{s})$ vanishes and one recovers the thin-shell Schwarzschild wormhole.

\subsection{Stable and unstable light rings}
The light rings correspond to photon orbits that satisfy $\dot{r}_\star=\ddot{r}_\star=0$. By applying these conditions to the geodesic equation~\eqref{xdot}, one obtains that these orbits obey
\begin{align}
\label{LR1}V_{\text{geo}}=E^2,\\
\label{LR2}\dfrac{d V_{\text{geo}}}{dr_\star}=0.
\end{align}
For light rays, instead of using $E$ and $L$, it is convenient to describe the orbits in terms of the impact parameter $b=L/E$. Therefore, in each side of the throat one can solve
\begin{align}
&\label{eq:b_c}b_c=\dfrac{r_\pm(r_{\star c})}{\sqrt{f_\pm(r_{\star c})}},\\
&\label{eq:r_c} r_\pm(r_{\star c})\dfrac{df_\pm}{dr_{\star}}\Big\vert_{r_{\star c}}-2f_\pm(r_{\star c})\dfrac{dr_\pm}{dr_\star}\Big\vert_{r_{\star c}}=0,
\end{align}
to find the light ring locations $r_{\star c}$ and their correspondent critical impact parameter $b_c$. As discussed, depending on the throat location, the dirty wormhole spacetime may inherit the unstable light rings from the  Schwarzschild spacetime, namely $r_\pm(r_{\star c})=3M$. If the matching surface is located precisely at $r_0=3M$, the two light rings merge in a single one. Using Eq.~\eqref{eq:b_c} one can find the critical impact parameter associated with these light rings, and compute the geometric absorption cross section of the thin-shell Schwarzschild wormhole, namely
\begin{equation}
\label{eq:abs_geo}\sigma_{geo} = \pi b_{c}^{2} = \begin{cases}27\pi M^2, &r_{0} < 3M, \\
\dfrac{r_{0}^{3}}{r_{0} - 2M}\pi, &r_{0} \geq 3M.
\end{cases}
\end{equation}
We note that the geometric absorption cross section depends on the radius of the throat for $r_0\geq 3M$, while as $r_{0} < 3M$, the absorption cross section is the same as the Schwarzschild black hole, regardless the throat location. This behavior of the absorption cross section can also be found in different wormhole configurations~\cite{Haroldo:2020}. In spherically symmetric spacetimes, the area obtained from Eq.~\eqref{eq:abs_geo} entirely determines the shadow of the compact object~\cite{Cunha:2018}, thus we note that thin-shell Schwarzschild wormholes may mimic some black hole properties. 

In the dirty wormhole case, Eq.~\eqref{eq:r_c} can also present additional three roots, $r_{+i}$ ($i=1,2$ and 3), besides the ones of the thin-shell wormhole without surrounding matter, namely
\begin{widetext}
\begin{align}
r_{+1} &= \dfrac{12r_s\Delta M(r_s+\Delta r_s)+\Delta r_s^3-\Delta r_s\sqrt{12\Delta M(r_s^2(3\Delta M + 2\Delta r_s)+2(r_s\Delta r_s(-3M+\Delta r_s)-3M\Delta r_s^2))+\Delta r_s^4}}{6\Delta M(2r_s+\Delta r_s)},\\
r_{+2} &= \dfrac{12r_s\Delta M(r_s+\Delta r_s)+\Delta r_s^3+\Delta r_s\sqrt{12\Delta M(r_s^2(3\Delta M + 2\Delta r_s)+2(r_s\Delta r_s(-3M+\Delta r_s)-3M\Delta r_s^2))+\Delta r_s^4}}{6\Delta M(2r_s+\Delta r_s)},\\
r_{+3} &= 3(M+\Delta M).
\end{align}
\end{widetext}
We point out that the roots $r_{+1}$ and $r_{+2}$ are found by considering the mass function $m(r_+)$ inside the thick shell, $r_s\leq r_+ \leq r_s+\Delta r_s$, while root $r_{+3}$ is found by considering $m(r_+)$ outside the thick shell, $r_+ > r_s+\Delta r_s$. Therefore, there are some constraints that roots must fulfill in order to actually represent extreme points of the effective potential, for instance
\begin{equation}
\label{eq:r12}r_s\leq r_{+1} \leq r_s+\Delta r_s, \,\,\text{and}\,\, r_s\leq r_{+2} \leq r_s+\Delta r_s,
\end{equation}
whereas
\begin{equation}
\label{eq:r3}r_{+3} \geq r_s+\Delta r_s.
\end{equation}

From conditions~\eqref{eq:r12} and~\eqref{eq:r3}, one notices that whether $\Delta M>0$, only the roots $r_{+2}$ and $r_{+3}$ represent extreme points of the effective potential for $r_+\geq r_s$. Additionally, $\Delta M\geq (r_s+\Delta r_s -3M)/3$ is required, for $r_{+2}$ to lie inside the thick shell and $r_{+3}$ to lie outside it. One can check, by computing $d^2V_{geo}/dr_{\star}^2$, that $r_{+2}$ corresponds to a local minimum, while $r_{+3}$ corresponds to a local maximum of the effective potential.

On the other hand, when $\Delta M<0$, only $r_{+1}$ and $r_{+2}$ are extreme points of the effective potential for $r_+\geq r_s$. The conditions~\eqref{eq:r12} and~\eqref{eq:r3} also impose a constraint on $\Delta M$, in order to $r_{+1}$ and $r_{+2}$ lie inside the thick shell, namely
\begin{equation}
\Delta M\leq -\dfrac{\Delta r_s}{6r_s^2}\left(A_s+\sqrt{\dfrac{B_s(3M-r_s)(A_s+r_s\Delta r_s)}{3M}}\right),
\end{equation}
where $A_s \equiv 2r_s(r_s+\Delta r_s)-3M(2r_s+\Delta r_s)$ and $B_s \equiv -3M(2r_s+\Delta r_s)$. An inspection on the behavior of $d^2V_{geo}/dr_{\star}^2$, shows that $r_{+1}$ corresponds to a local maximum, while $r_{+2}$ corresponds to a local minimum of the effective potential. A careful look on the roots $r_{+1}$ and $r_{+2}$, shows that they can be degenerate for a certain value of $\Delta M$. However, when it happens, $d^2V_{geo}/dr_{\star}^2=0$, and one cannot infer if it corresponds to a maximum or a minimum of the effective potential.

Therefore, we notice that whenever the model~\eqref{mass_profile} introduces new light rings in the spacetime, it introduces them in pairs, one being stable and another one being unstable. This property is related to the conservation of the total topological charge associated to the number of light rings in the wormhole spacetime~\cite{xavier2024traversable} (see also Ref.~\cite{Cunha:2017, Cunha:2020}).

When more than one maximum is present in the effective potential of a spherically symmetric spacetime, its shadow is associated to the dominant light ring (the highest peak in the effective potential)~\cite{Merce:2021, junior2021can}. Since new maxima are introduced in the effective potential of a dirty wormhole, the dominant light ring may change. It happens below a certain value of $\Delta M$, namely $\Delta M_c$, which we do not exhibit here since it is a cumbersome function of the model parameters. Remarkably, $\Delta M_c$ is independent of the throat location if $r_0\leq 3M$, but dependent on it if $r_0>3M$.  One can see this behavior in the bottom panel of top-left quadrant of Fig.~\ref{VgeoM}, in which the narrower peak is higher than the peaks inherited from the Schwarzschild spacetime. Therefore the shadow of this spacetime departs from the case without surrounding matter.

\subsection{Photon orbits}
As mentioned above, it is convenient to study the photon orbits in terms of the impact parameter $b$ rather than using $E$ and $L$. Additionally, it is also suitable to define $u_{\pm}\equiv 1/r_{\pm}$, for expressing the orbit equation on each side of the throat, that is
\begin{equation}
\dfrac{d^2u_{\pm}}{d\phi^2} = -\dfrac{1}{2}\left[2f_{\pm} u_\pm+\dfrac{df_{\pm}}{du_{\pm}}u_{\pm}^2\right].
\end{equation}
To obtain the null orbits impinging from one side, one starts from an initial angle $\phi_0$ and evolves the orbit equation considering the conditions
\begin{equation}
u_{\pm}(\phi_0) = u_0 \,\,\,\text{and}\,\,\, \left(\dfrac{du_\pm}{d\phi}\right)^2\bigg\vert_{\phi_0} = \left(\dfrac{1}{b^2}-u_\pm^2f_{\pm}\right)\bigg\vert_{\phi_0},
\end{equation} 
depending on the side one chooses to start the integration.

\begin{figure*}[!h]
\includegraphics[width=\columnwidth]{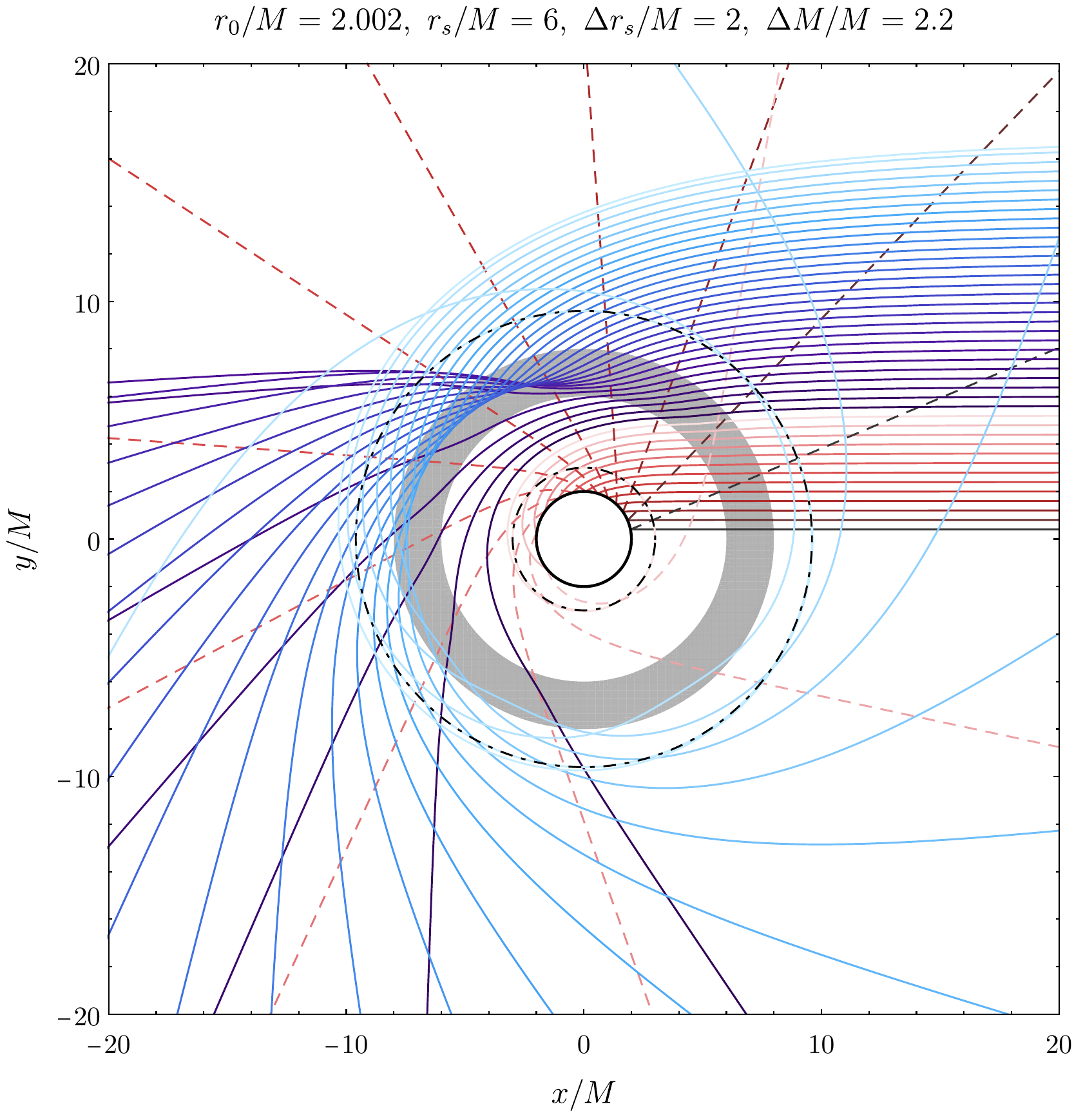}\includegraphics[width=\columnwidth]{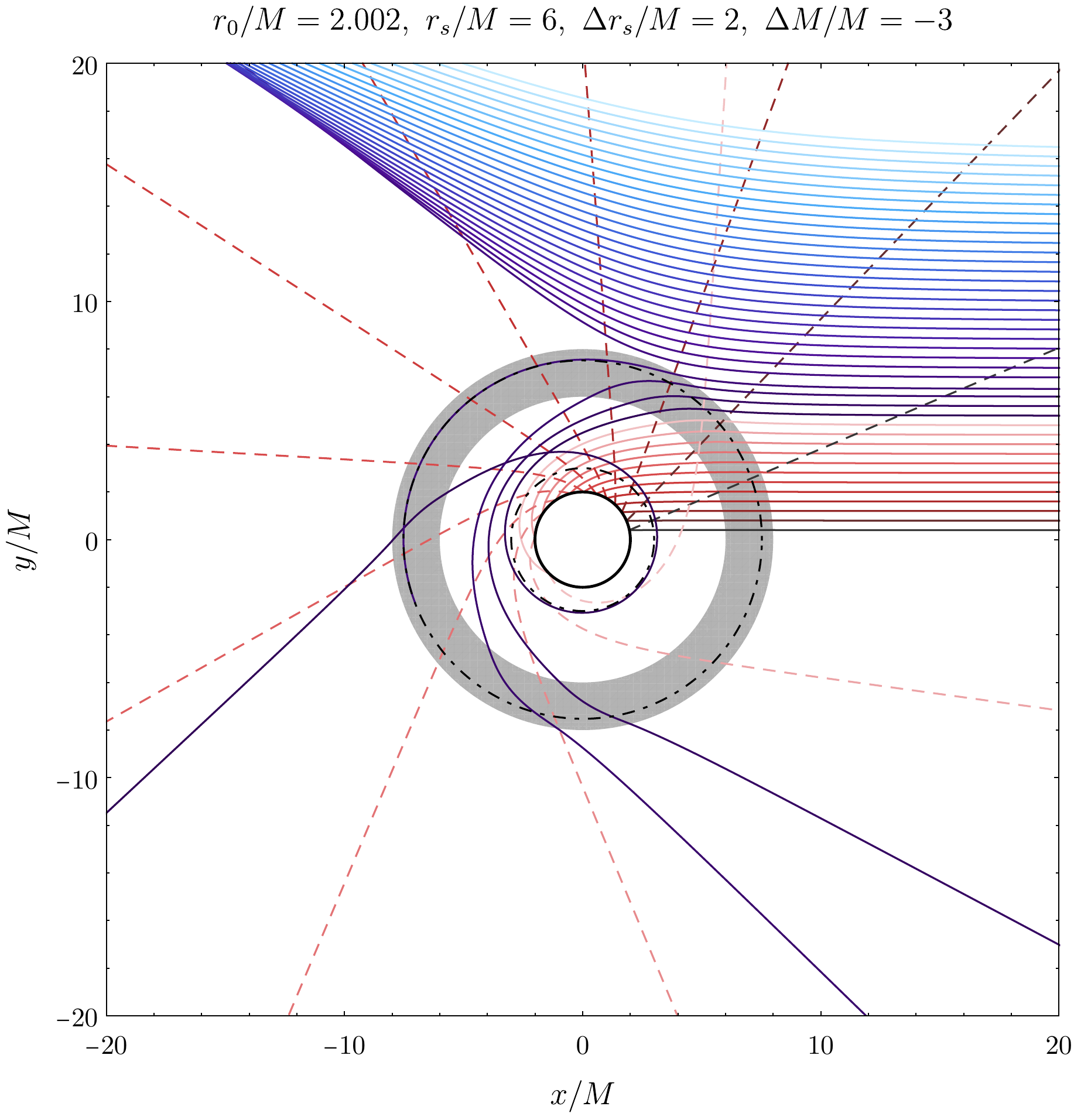}\\
\includegraphics[width=\columnwidth]{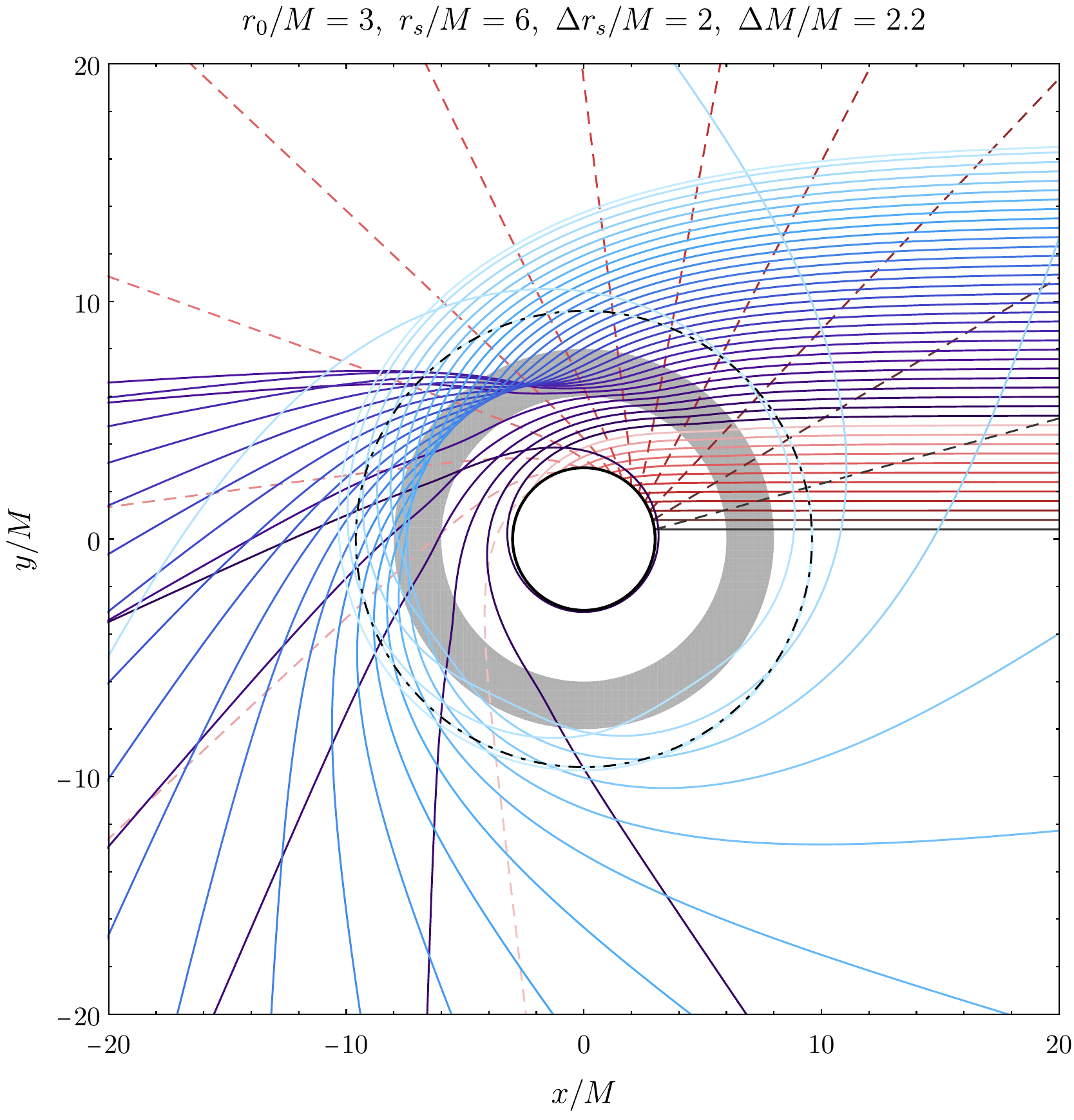}\includegraphics[width=\columnwidth]{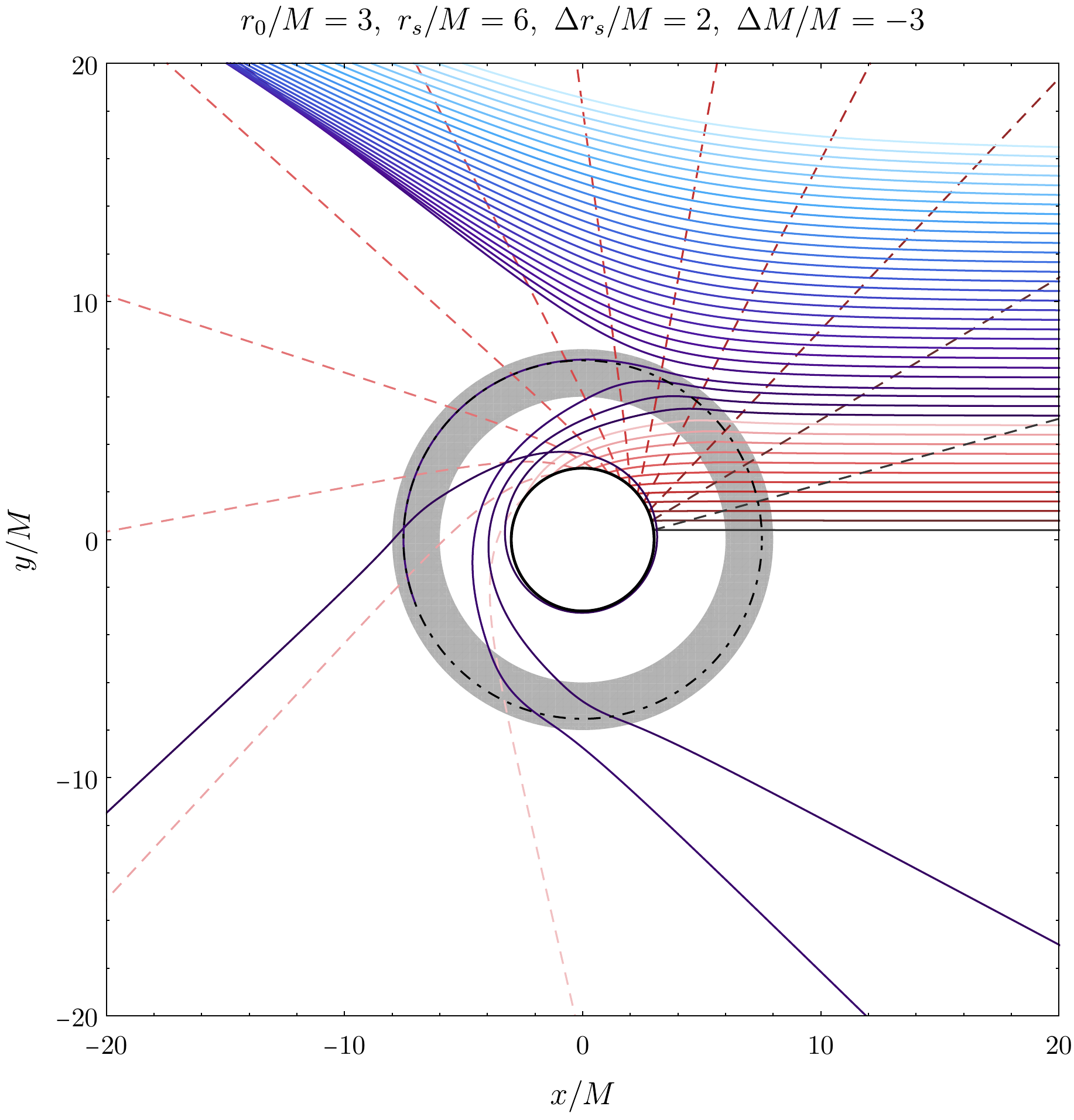}
\caption{Bending of light in dirty wormhole backgrounds. The dirtiness is represented by the gray strip and we are considering two values of the thick shell mass, namely $\Delta M=2.2M$ and $\Delta M=-3M$. Both cases introduce new light rings in the spacetime. In the positive thick shell mass case, the new unstable light ring is located outside the shell, while in the negative thick shell mass case, the new unstable light ring is located inside the thick shell. The unstable light rings are represented by dot-dashed circles, including the innermost ones at $r_\pm = 3M$, inherited from the thin-shell surgery. In the top row, the throat is located at $r_0=2.002M$, while in the bottom row the throat is located at $r_0=3M$, and we represent the throat by the solid circle. The light rays colored by the blue gradient represent geodesics scattered by the dirty wormhole (geodesics that do not cross the throat), while the light rays colored by the red gradient depict curves absorbed by the wormhole (geodesics that cross the throat). After crossing the throat, we choose dashed lines to represent the curves in the inner universe.}
\label{fig:orbits}
\end{figure*}

In Fig.~\ref{fig:orbits}, we exhibit a selection of null orbits in the dirty wormhole background, considering thick shells distributed between $6M$ and $8M$, and with two values of mass, namely $\Delta M =2.2M$ and $\Delta M =-3M$. We also considered two different throat locations, that is $r_0=2.002M$ and $r_0=3M$. A glance at Fig.~\ref{VgeoM} shows that these two dirtiness profiles introduce additional peaks in the effective potential, related with new unstable light rings in the geometry. For $\Delta M > 0$, the peaks inherited from the thin-shell surgery are still related to the dominant light ring, and therefore the shadow of these configurations remains the same. 

From Fig.~\ref{fig:orbits} one notices that matter in the environment of wormholes may sharply influence the scattering of light rays as much as more mass is considered.
By considering matter with negative energy density (phantom matter), if $\Delta M<-M$, the spacetime outside the shell corresponds to a Schwarzschild spacetime with negative mass, therefore one expects repulsion effects when approaching the matter shell. One notices this behavior in the right column of Fig.~\ref{fig:orbits}, where, for certain values of the impact parameters, the light rays are repelled by the matter distribution. In particular, for the phantom matter distribution considered in Fig.~\ref{fig:orbits}, null geodesics with $b\approx 6.31929M$ stay trapped inside the matter shell, while the geodesics entering the matter shell with impact $3\sqrt{3}M<b < 6.31929M$ are scattered back to infinity of ${\cal M}_+$. Light rays with impact parameter $b<3\sqrt{3}M$ cross the throat and propagate toward the infinity of ${\cal M}_-$.

By considering matter shells with positive energy density (normal matter), the spacetime outside the thick shell is described by a Schwarzschild spacetime with positive net mass $M+\Delta M$, and the behavior of light rays outside the shell is what one expects from photon trajectories outside a Schwarzschild black hole with mass $M+\Delta M$. 
Some of these light rays cannot penetrate entirely the shell and are scattered back to the infinity of $\mathcal{M}_{+}$. This behavior can be visualized in the left column of Fig.~\ref{fig:orbits}, where one notices some geodesics being bent inside the shell such that they propagate outside it. From the effective potential perspective, this behavior corresponds to the light rays with values of impact parameter such that they face the potential barrier and therefore reach a turning point at $r_s<r_b<r_s+\Delta r_s$. One can compute the value of the impact parameter of a light ray entering the shell, reaching a turning point at the innermost edge of the thick shell, and being scattered back to infinity, namely $b=r_s/\sqrt{f_{+}(r_s)}$. For such impact parameter, the turning point of the light ray coincides with $r_s$. Therefore, for the matter distribution considered in the left column of Fig.~\ref{fig:orbits}, any photon with impact parameter $b>3\sqrt{6}M$ cannot penetrate the shell. Light rays with impact parameter $3\sqrt{3}M<b<3\sqrt{6}M$ penetrate the shell but are scattered back to the infinity of ${\cal M}_+$, and light rays with impact parameter $b<3\sqrt{3}M$ do cross the throat and propagate toward the infinity of ${\cal M}_-$.

Before we proceed to the next section, let us discuss the orbits in a peculiar configuration of a wormhole surrounded by a phantom matter shell with mass $\Delta M=-M$. In this configuration, outside the thick shell, the spacetime corresponds to a Minkowski spacetime, since $M+\Delta M=0$ above $r_s+\Delta r_s$. Hence, light rays that do not encounter the thick shell, propagate as straight lines without being bent. Null geodesics with impact parameter $b<r_s+\Delta r_s$ suffer a deflection in their trajectories and can stay trapped either in an unstable light ring within the thick shell or in an unstable light ring below it, inherited from the thin-shell surgery. In Fig.~\ref{fig:mink_outside}, we exhibit some photon orbits around this dirty wormhole configuration with $\Delta M=-M$, distributed between $6M$ and $8M$. 

\begin{figure}[!h]
\includegraphics[width=\columnwidth]{sctPM-2-0M--1-0M-duplo.pdf}
\caption{Photon orbits in the outside (${\cal M}_+$) of a wormhole surrounded by a matter shell with $\Delta M/M=-1$. Null geodesics with $b>r_s+\Delta r_s$ do not feel any gravitational attraction or repulsion, therefore propagate as straight lines. Light rays that encounter the shell (gray strip) are bent and can be scattered back to the infinity of ${\cal M}_+$ (lines colored by blue gradient) or absorbed by the wormhole throat (lines colored by red gradient). The wormhole throat and the unstable photon spheres are represented by a solid circle and a dot-dashed circle, respectively.}
\label{fig:mink_outside}
\end{figure}

Our results show that the trajectory of photons around wormholes may be sharply influenced by their environment. This indicates that matter surrounding wormholes may produce remarkable gravitational lensing effects even though the shadow of the wormhole remains the same.

\section{Quasibound states and spectral lines}\label{sec:abs}
When ECOs with stable photon sphere are perturbed, quasitrapped modes are expected to exist~\cite{Caio:2018}. These modes may give rise to a series of echoes in the ringdown profile at late times~\cite{KZ:2019}, what could be used to tell these ECOs apart from black holes and neutron stars, for example. Another remarkable consequence of the existence of quasibound states is the appearance of resonant peaks in the absorption spectra~\cite{Caio:2018,Haroldo:2020}, that come as sharp peaks in the transmission coefficients and in the absorption cross section, what introduces notable differences in the absorption bands, when compared to standard black holes. 

As discussed in the previous section, the presence of matter in the neighborhood of wormholes may introduce new photon spheres in the spacetime structure, despite the ones inherited from the matching of two Schwarzschild spacetimes. When these new light rings appear they come in pairs, one of them being stable~\cite{xavier2024traversable}. Therefore, additional quasibound states may exist near the matter shell, what could affect the echoes profile. In Ref.~\cite{KZ:2019} was studied the influence of the thick matter shell near wormholes in the time profile of scalar field perturbations, where the authors found that shells with small masses do not produce significant effects in the echoes profile, and only heavier configurations could remarkably affect it. In order to complement this analysis, let us investigate how the surrounding matter influences the absorption of scalar waves, and see how the environment affects the absorption bands.

\subsection{Scalar field dynamics}
Let us consider a test massless scalar field, $\Phi$, lying in the vicinity of a dirty wormhole. In each side of the throat, the dynamics of the field is encoded in the correspondent Klein-Gordon equation, namely
\begin{equation}
\label{eq: MKG} \Box_{\pm} \Phi_{\pm} = 0,
\end{equation}
where $\Box_{\pm}$ and $\Phi_{\pm}$ are the wave operator and the scalar field, respectively, on each side of the throat. The field is assumed to be continuous across the thin shell, therefore $[\Phi]=0$, and since the metric and its determinant are also continuous across the shell, no delta-type distribution will appear in the effective potential, thus no discontinuity on the derivatives of the field are expected~\cite{magalhaes2023asymmetric}.

Due to the spherical symmetry, the scalar field can be cast as
\begin{equation}
\label{eq: Scalar} \Phi(t,r_{\pm},\theta,\phi) = \frac{\psi_\pm(r_\pm)}{r_\pm}Y_{\ell m}e^{-i\omega t},
\end{equation}
where $Y_{\ell m}$ is the spherical harmonic with multipole $\ell$, azimutal number $m$ and $\omega$ is the frequency of the scalar wave. 
The radial functions $\psi_\pm$, satisfy on each side of the throat,
\begin{equation}
f_{\pm}(r_\pm)\dfrac{d}{dr_\pm}\left(f_\pm(r_\pm)\dfrac{d\psi_\pm}{dr_\pm}\right)+(\omega^2-V_\pm(r_\pm;\ell))\psi_\pm = 0,
\end{equation}
where
\begin{equation}
\label{eq: Veff} V_{\pm}(r_\pm;\ell) = f_\pm(r_\pm)\left(\frac{1}{r_\pm}\frac{df_\pm}{dr_\pm} + \frac{\ell(\ell+1)}{r_\pm^{2}} \right)
\end{equation}
is the effective potential on ${\cal M}_{\pm}$. In terms of the global coordinate, the radial function satisfies a Schr{\"o}dinger-like equation, namely
\begin{equation}
\label{eq: ERG} \frac{d^{2}\psi}{dr_\star^{2}} + \left[ \omega^{2} - V(r_\star;\ell)\right]\psi= 0,
\end{equation}
where we have dropped the $\pm$ subscript, since the radial function and the effective potential depend solely on the global coordinate.

\begin{figure*}
\includegraphics[width=\columnwidth]{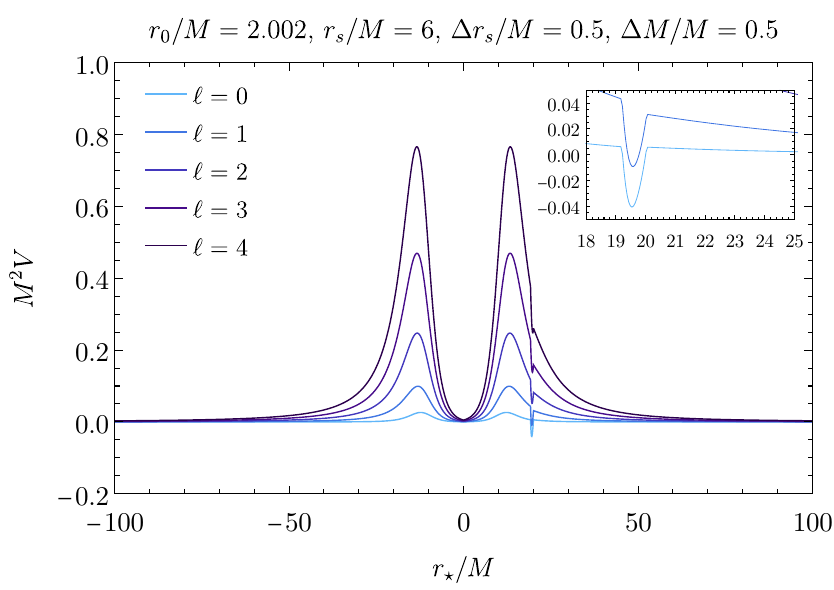}\includegraphics[width=\columnwidth]{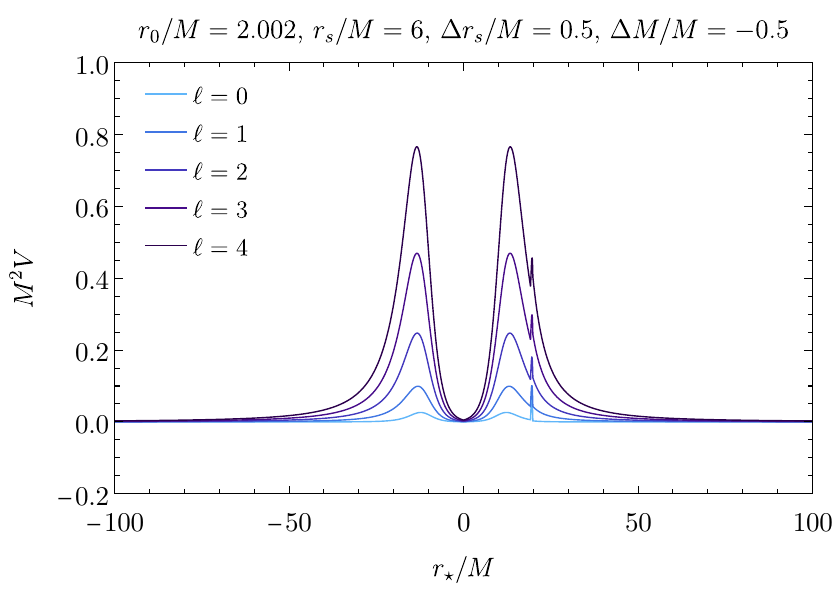}\\
\includegraphics[width=\columnwidth]{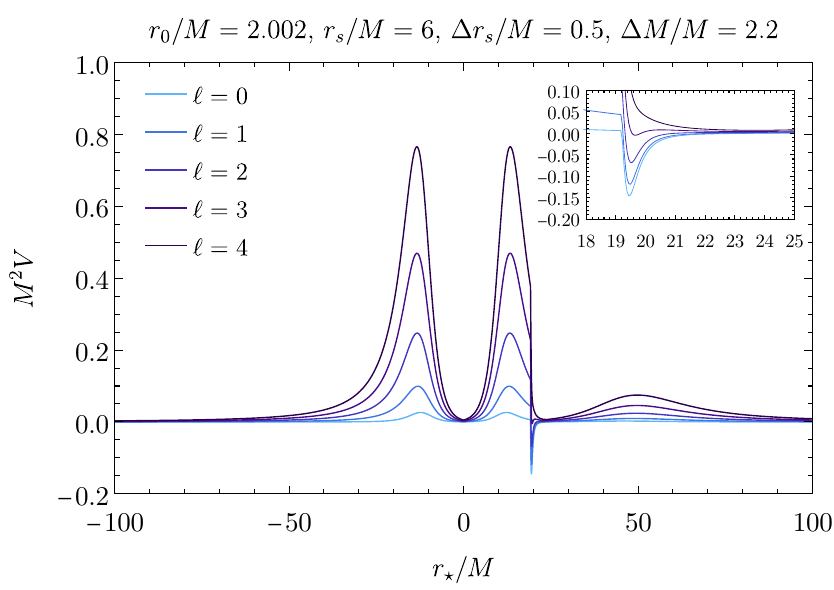}\includegraphics[width=\columnwidth]{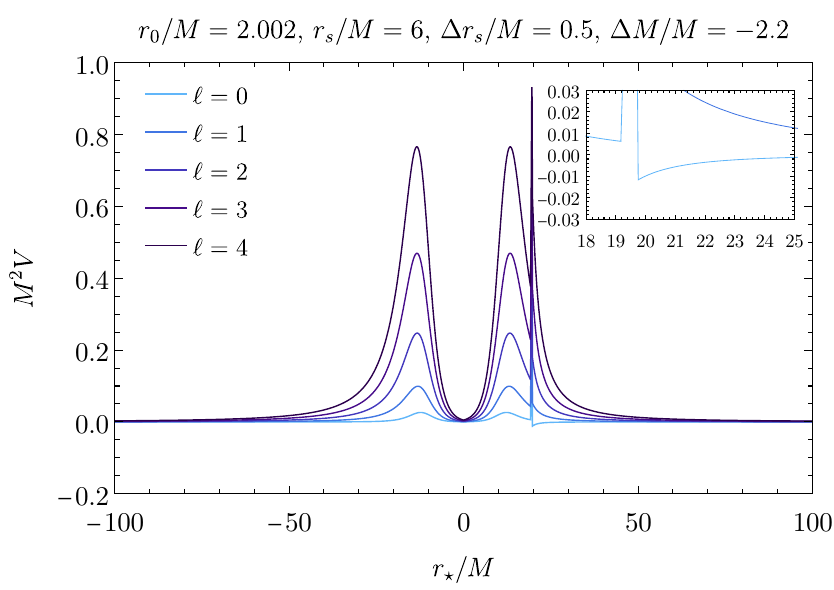}
\caption{Massless scalar field's effective potential of some dirty wormhole configurations, for different multipole numbers $\ell$. Contrasting with the geodesic effective potential, depending on the thick shell distribution, regions of negative effective potential are present for small multipole numbers.}
\label{fig:eff_pot_field}
\end{figure*}

In Fig.~\ref{fig:eff_pot_field} we show the effective potential of the scalar field, for some dirty wormhole configurations with some multipole numbers $\ell$. For higher multipole modes, the effective potential goes to the shape of the null geodesic's effective potential. For the cases exhibited in Fig.~\ref{fig:eff_pot_field}, the wormhole throat is located inside the Schwarzschild's photon sphere. Therefore, a stable photon sphere is present, which leads to a valley in the effective potential at the throat. When the scalar wave interacts with the wormhole, some quasibound states arise near the throat. These modes are allowed to tunnel and can be measured, for example, as a series of echoes in the time profile of the scalar wave. Before further discussion on these modes and how to compute them, let us focus on how the surrounding matter influences the effective potential of the scalar field.

As in the case of null geodesics, depending on the distribution, phantom matter surrounding the wormhole may introduce sharp peaks in the effective potential, while normal matter in the wormhole environment may create new valleys in the effective potential profile.  In the top row of Fig.~\ref{fig:eff_pot_field}, we exhibit thick shells with mass less than the wormhole, while in the bottom row we consider heavier thick shells. Even though the latter configurations represent less astrophysical environments, they highlight an interesting behavior, that is, the existence of regions with negative effective potential well for lower multipole number $\ell$. Remarkably, this behavior persists even for not so heavy configurations.

The new valleys and peaks, introduced due to the presence of surrounding matter, may change the number, frequency and damping of the quasibound modes. These differences in the quasibound states can introduce deviations in the absorption bands, for instance the spectral lines in the absorption profile may suffer a shift or even new resonant peaks may appear.
\subsection{Quasibound states} 
As previously pointed out, the presence of a well in the effective potential~\eqref{eq: Veff} allows quasibound states of the scalar field to arise near the stable photon spheres.  
These modes are characterized by complex frequencies with small imaginary part, which are solutions of an eigenvalue problem to $\omega$, when the boundary conditions
\begin{equation}
\label{eq: BCQBS}  \psi(r_\star) \approx \begin{cases} 
e^{i\omega r_\star}, &r_\star \rightarrow +\infty,\\
e^{-i \omega r_\star},&r_\star\rightarrow -\infty,
\end{cases}
\end{equation}
are considered.
Equation~\eqref{eq: BCQBS} represents purely outgoing scalar waves at both sides of the dirty wormhole spacetime. In order to determine those complex frequencies, we employ the direct integration method~\citep{Chandra:1975}. The method consists in numerically integrating the Eq. \eqref{eq: ERG} from the asymptotic regions to the same intermediary point $r_{\star m}$, at which one calculates the Wronskian of the two numerical solutions and applies a root-finding method to obtain a frequency $\omega$, such that the Wronskian is zero. This method is independent of the intermediary point~\citep{Pani:2013}. 

In Table~\ref{tab:Clean_TS}, we exhibit the real and imaginary parts, respectively, $\omega_r$ and $\omega_i$, of the complex frequencies of some quasibound states present near the throat of a thin-shell Schwarzschild wormhole without surrounding matter. Those are long lived modes, since they have small imaginary part. The modes shown in Table~\ref{tab:Clean_TS} correspond to the numerical solutions with good convergence, that remain stable as we increase the numerical infinity. We also found other quasibound states for the considered multipole modes, however they correspond to numerical solutions with poorer convergence, and one should enhance the numerical method to guarantee their existence and precision. This could be done, for instance, by considering an expansion at both infinities as done in~\cite{Cardoso:2016}. However, the modes present here (and hereafter) are enough to illustrate the effect of trapped modes on the absorption phenomena.
	\begin{table}[h!]
		\centering \caption{Quasibound frequencies of a thin-shell Schwarzschild wormhole with throat located at $r_0 = 2.002M$.}
		\vskip 10pt
		\begin{tabular}{@{}ccccccc@{}}
			\hline \hline
			%
			$\ell$\hspace{0.5cm}      &$M\omega_r$\hspace{0.5cm} &$-M\omega_i$ \\
\hline \hline       
$0$\hspace{0.5cm}            &$ 0.1055 $\hspace{0.5cm}         		& $7.25\times 10^{-3}$ \\
			\hline
			$1$\hspace{0.5cm}               &$ 0.1481  $\hspace{0.5cm}          &$9.71\times 10^{-5} $        \\
			\hspace{0.5cm}               &$ 0.2608  $\hspace{0.5cm}          &$ 4.04\times 10^{-3} $          \\
			\hline
			$2$\hspace{0.5cm}               &$ 0.1807  $\hspace{0.5cm}          &$  2.34\times 10^{-7} $        \\
			\hspace{0.5cm}               &$ 0.3205  $\hspace{0.5cm}          &$ 6.28\times 10^{-5} $        \\
			\hspace{0.5cm}               &$ 0.4284  $\hspace{0.5cm}          &$ 2.08\times 10^{-3} $       \\
			\hline \hline
		\end{tabular}
		\label{tab:Clean_TS}
	\end{table} 

In Table. \ref{tab:dirty_WH} we exhibit the real and imaginary parts of the frequencies associated with some quasibound states present in two dirty wormholes spacetimes. In both configurations, the thick shell is located between $6M$ and $6.5M$. For the phantom matter case, we considered a shell with $\Delta M = -3M$, while for the normal matter configuration, the shell has mass of $\Delta M = 2.2M$. One notices that the presence of matter in the environment can modify the set of quasibound states, when compared to the thin-shell wormhole without surrounding matter. 
As pointed out in Ref.~\cite{KZ:2019}, these deviations in the quasibound mode spectra are expected to be more relevant for heavier thick-shells configurations. The dirty wormholes considered in this table, even though less astrophysical relevant, reveal an interesting behavior. One notices that, even for heavy distributions, the presence of matter around wormholes does not change much the real part of the quasibound frequencies of preexistent modes that also appear when no shell of matter is considered. In particular, one notices a very good agreement between the real parts of the quasibound frequencies of wormholes with and without surrounding matter, for $\ell=1$ and $\ell=2$. On the other hand, one notices that the imaginary parts of the quasibound states in dirty wormholes, really depart from the case without surrounding matter, and therefore the damping of these modes can be very different. 
	\begin{table}[hbtp!]
		\centering \caption{Quasibound frequencies of dirty wormholes with throat located at $r_0 = 2.002M$, and thick shells of matter located between $6M$ and $6.5M$.}
		\vskip 10pt
		\begin{tabular}{@{}ccc|ccc@{}}
			\hline \hline
			\multicolumn{3}{c|}{$\Delta M/M=-3$} & \multicolumn{3}{c}{$\Delta M/M=2.2$}\\\hline\hline
			$\ell$\hspace{0.5cm}      &$M\omega_r$\hspace{0.5cm} &$-M\omega_i$ &$\ell$\hspace{0.5cm}      &$M\omega_r$\hspace{0.5cm} &$-M\omega_i$ \\
\hline \hline       
$0$\hspace{0.5cm}            &$ 0.1071 $\hspace{0.5cm}         		& $5.21\times 10^{-3}$  &$0$\hspace{0.5cm}            &$ 0.0407 $\hspace{0.5cm}         		& $4.52\times 10^{-3}$ \\\hspace{0.5cm}            &\hspace{0.5cm}         		&  &\hspace{0.5cm}            &$ 0.1099 $\hspace{0.5cm}         		& $9.45\times 10^{-3}$ \\
			\hline
			$1$\hspace{0.5cm}               &$ 0.1481  $\hspace{0.5cm}          &$7.38\times 10^{-5} $        &$1$\hspace{0.5cm}               &$ 0.0901  $\hspace{0.5cm}          &$2.92\times 10^{-3} $
			\\
			\hspace{0.5cm}               &$ 0.2609  $\hspace{0.5cm}          &$ 2.81\times 10^{-3} $          & 
			\hspace{0.5cm}               &$ 0.1480  $\hspace{0.5cm}          &$ 3.27\times 10^{-4}  $\\
			\hspace{0.5cm}            &\hspace{0.5cm}         		&  &\hspace{0.5cm}            &$ 0.2621 $\hspace{0.5cm}         		& $3.99\times 10^{-3}$ \\
			\hline
			$2$\hspace{0.5cm}               &$ 0.1807  $\hspace{0.5cm}          &$  2.01\times 10^{-7} $       &$2$\hspace{0.5cm}               &$ 0.1253   $\hspace{0.5cm}          &$  3.51\times 10^{-4} $       \\
			\hspace{0.5cm}               &$ 0.3205  $\hspace{0.5cm}          &$ 4.12\times 10^{-5} $        &
			\hspace{0.5cm}               &$ 0.1807  $\hspace{0.5cm}          &$ 3.76\times 10^{-6}  $        \\
			\hspace{0.5cm}               &$ 0.4281  $\hspace{0.5cm}          &$ 1.79\times 10^{-3} $       &\hspace{0.5cm}               &$0.3205  $\hspace{0.5cm}          &$ 9.29\times 10^{-5}$       \\			%
			\hspace{0.5cm}            &\hspace{0.5cm}         		&  &\hspace{0.5cm}            &$ 0.4288 $\hspace{0.5cm}         		& $1.80\times 10^{-3}$ \\

			\hline \hline
		\end{tabular}
		\label{tab:dirty_WH}
	\end{table}  
	
Remarkably, the presence of heavier distributions can also introduce additional quasibound frequencies in the spectrum, as noticed in the case of thick shell with $\Delta M=2.2M$. These new quasibound frequencies are related to the stable photon sphere that arise due to the presence of a thick shell. As a consequence, the presence of matter in the environment of compact objects can confine modes in the surrounding of the central object. 
\subsection{Absorption bands}
Since the spectra of quasibound states of dirty wormholes deviate from the case without surrounding matter, one expects absorption bands of dirty wormholes with shifted and possibly new resonant peaks, presenting thus a different pattern of spectral lines. 

We analyze scalar waves subjected to scattering boundary conditions, that is, the radial function must satisfy, asymptotically,
\begin{equation}
\label{eq:BC_scattering} \psi(r_\star) \approx \begin{cases} 
e^{-i\omega r_\star} + R_{\omega \ell}e^{i \omega r_\star}, &r_\star \rightarrow +\infty,\\
T_{\omega \ell}e^{-i\omega r_\star} ,&r_\star\rightarrow -\infty,
\end{cases}
\end{equation}
where $R_{\omega \ell}$ and $T_{\omega \ell}$ are, respectively, associated to the reflection and transmission coefficients.
These boundary conditions are associated to a monochromatic plane wave impinging from the past null infinity of ${\cal M}_+$, interacting with the dirty wormhole, and being partially transmitted to the  future null infinity of ${\cal M}_-$, and partially scattered back to the future
null infinity of ${\cal M}_+$. That is, far from the object, in one side of throat, there is a composition of ingoing and outgoing distorted planes waves, and purely outgoing waves on the other side of the throat. 

By considering the boundary conditions~\eqref{eq:BC_scattering} and a plane wave expansion, one can compute the total scalar absorption cross section as a sum over the partial absorption cross sections, $\sigma_\ell$, namely
\begin{equation}
\label{eq: Sabs} \sigma_\text{abs} = \sum_{\ell = 0}^{\infty}\sigma_\ell,
\end{equation}
where $\sigma_\ell \equiv \pi (2\ell+1)|T_{\omega \ell}|^2/\omega^2$. The so-called transmission coefficients $|T_{\omega \ell}|^2$ give the transmission probability of a mode with frequency $\omega$ and multipole $\ell$, that from the flux conservation law satisfies $|T_{\omega \ell}|^2+|R_{\omega \ell}|^2=1$, with $R_{\omega \ell}|^2$ being the reflection coefficients. To obtain $|T_{\omega \ell}|^{2}$, we integrate numerically the radial equation~\eqref{eq: ERG} from one asymptotic region to the other, namely from $r_\star\to -\infty$ to $r_\star\to +\infty$, and then we compare the numerical solution with the boundary condition~\eqref{eq:BC_scattering} at $r_\star\to +\infty$.

It is important to point out that the total scalar absorption cross section has two well-known limits for stationary black hole holes without surrounding matter. In the high-frequency regime, one can approximate the absorption process using geodesic motion~\cite{Decanini:2011}, and one finds that the total absorption cross section oscillates around the geometrical absorption cross section, according to the so-called sinc approximation~\cite{Sanchez:1978}. On the other hand, in the low-frequency regime, $\sigma_\text{abs}$ is dominated by the lowest multipole number $\ell=0$~\cite{Unruh:1976}, and in this limit the absorption cross section goes to the area of the event horizon of stationary black holes~\cite{Higuchi:2001}. These two limits are generic results for stationary black holes in electrovacum, and also tested in several other scenarios~\cite{Caio:2018,Haroldo:2020,magalhaes2023asymmetric}. 

For wormhole scenarios these behaviors are more subtle. As investigated in previous works, the eikonal limit of the total absorption cross section of wormholes also oscillates around the geometrical absorption cross section. In particular, in asymmetric scenarios, where two photon spheres with different radii are present, the scalar absorption cross section oscillates around the geometrical absorption cross section associated to the dominant light ring~\cite{magalhaes2023asymmetric}. On the other hand, the low-frequency limit of the wormhole's absorption process was not investigated in details yet, and further analysis should be done in order to understand this regime.

\subsection{Numerical results}
\subsubsection{Transmission coefficients}
\begin{figure}[!h]
\includegraphics[width=\columnwidth]{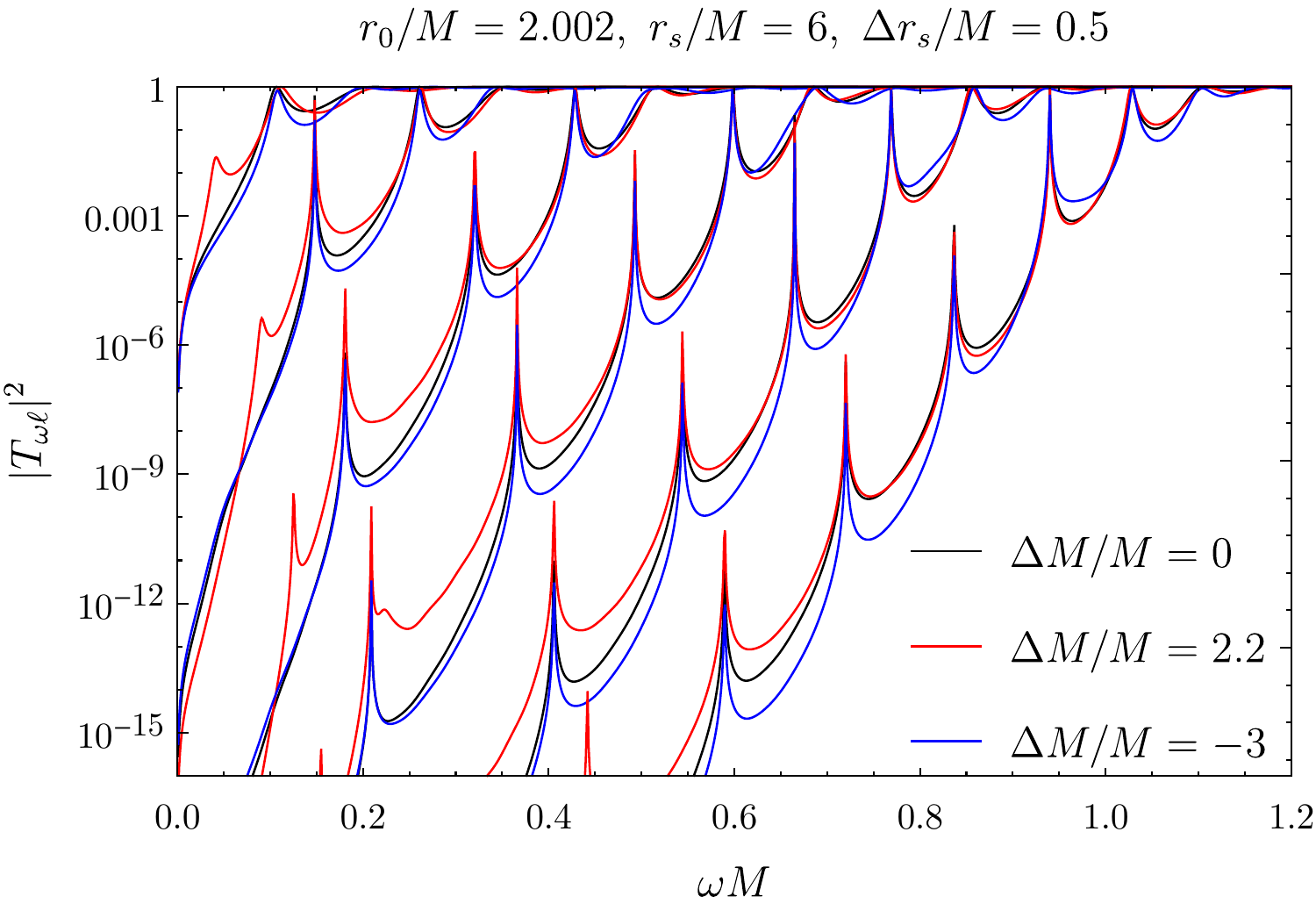}
\caption{Transmission coefficients of wormholes with and without surrounding matter. The configurations considered in this figure are the same as the ones considered in Tables~\ref{tab:Clean_TS} and~\ref{tab:dirty_WH}. One notices that the number of trapped modes (and consequently resonant peaks) increases as larger multipole numbers are considered. In this figure we considered from $\ell=0$ to $\ell=5$.}
\label{fig:tra_coef}
\end{figure}

In Fig.~\ref{fig:tra_coef} we show the transmission coefficients of the dirty wormholes considered in Table~\ref{tab:dirty_WH} and compare them with the case without surrounding matter represented in Table~\ref{tab:Clean_TS}. The remarkable aspect of these transmission coefficients is the presence of narrow peaks at specific frequency values. One notices that these peaks are located at the real part of the quasibound state frequencies. Using the Breit-Wigner expression for nuclear scattering, one can approximate the transmission probability by~\cite{breit1936capture}:
\begin{equation}
\label{eq:BW} |T_{\omega \ell}|^2=\dfrac{A_{\omega\ell}}{(\omega-\omega_r)^2+\omega_i^2},
\end{equation}
where $A_{\omega\ell}$ are constants evaluated near each peak, that depend on the frequency and on the multipole number. From this expression, one notices that the narrow peaks are centered at the real part of the quasibound frequency, and the imaginary part determines the sharp shape and height of the peak.

\begin{figure*}
\includegraphics[width=\columnwidth]{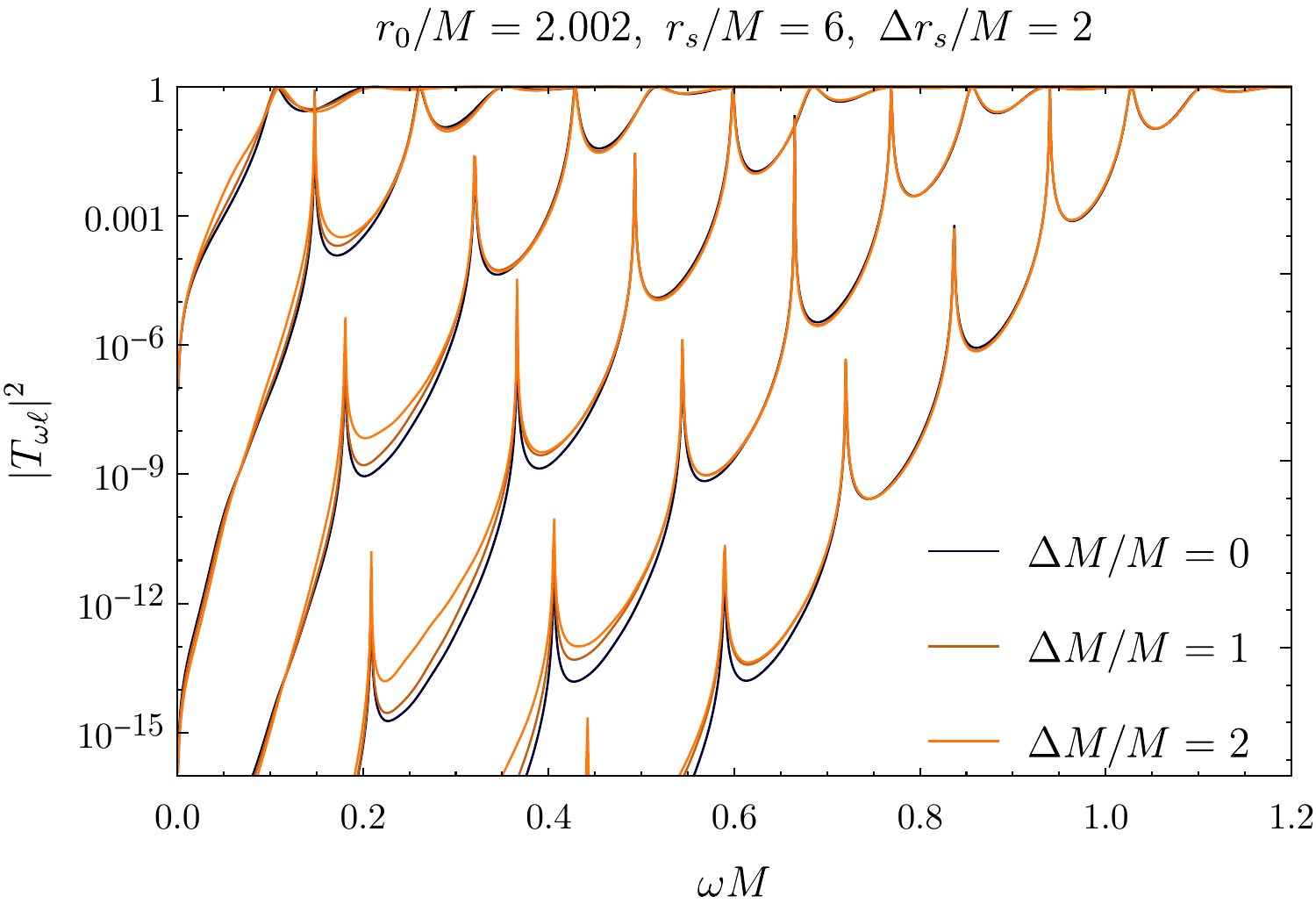}\includegraphics[width=\columnwidth]{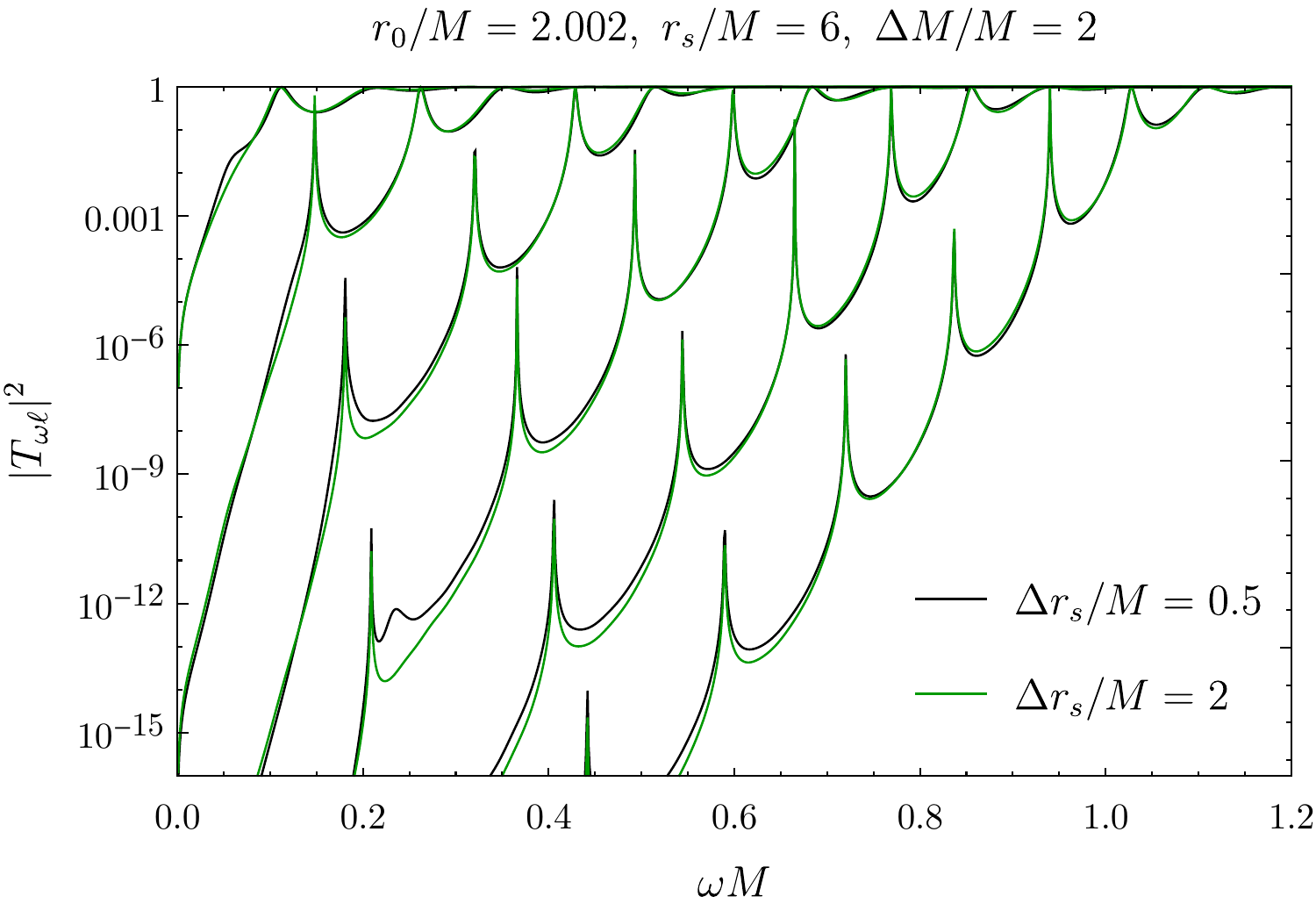}\\
\includegraphics[width=\columnwidth]{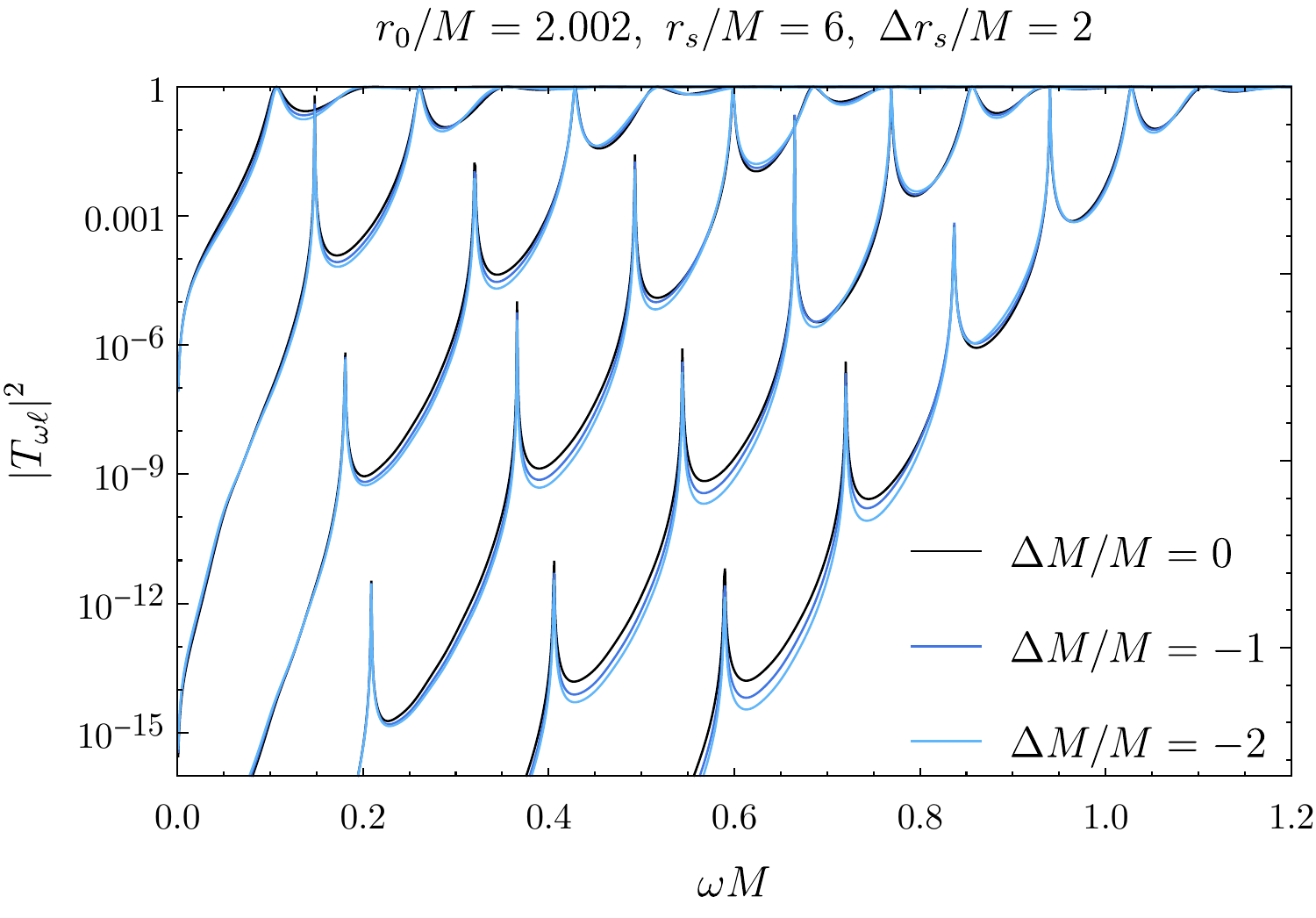}\includegraphics[width=\columnwidth]{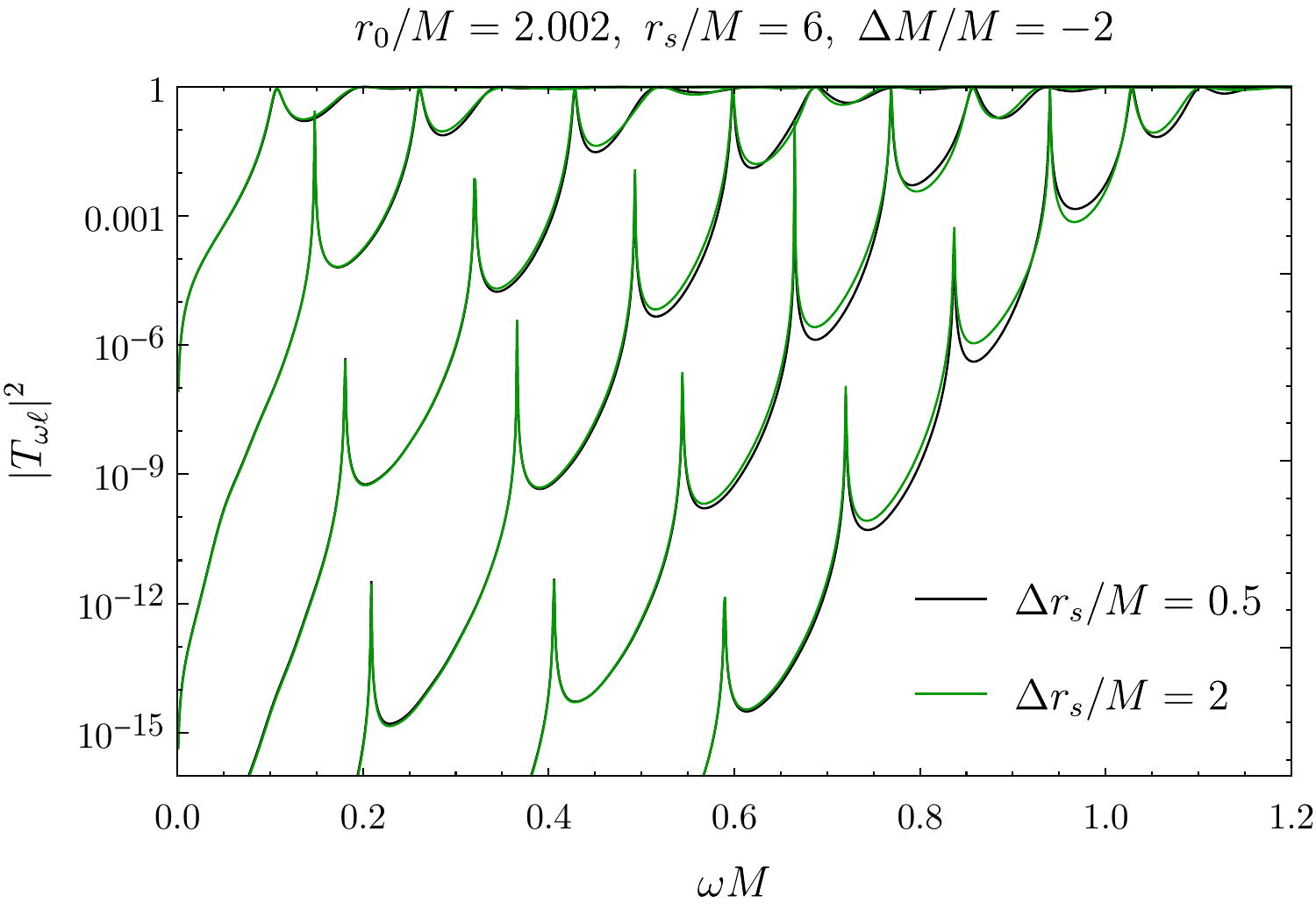}
\caption{Effect of the matter shell mass and thickness on the amplitude of the transmission probability. In the left column panels we show how different values of thick-shell mass modify the transmission coefficients. In the right column panels we show the influence of the thickness on the transmission coefficients. In this figure we considered from $\ell=0$ to $\ell=5$.}
\label{fig:tra_coef_thickness}
\end{figure*}

From Fig.~\ref{fig:tra_coef} it is evident that the surrounding matter can modify the amplitude of the transmission probability, in particular introducing new narrow peaks in the transmission coefficients. This is related to new trapped modes found when considering matter around the wormhole throat. To have a better understanding of how the matter distribution surrounding the wormhole can affect the transmission probability, we plotted in Fig.~\ref{fig:tra_coef_thickness} the transmission coefficients for different thick-shell masses and thickness. Our results indicate that the environment can indeed modify the amplitudes of the transmission coefficients, however some fingerprints of the isolated wormhole are kept. Specifically, one notices that the resonant peaks present in the case without thick shell of matter surrounding the wormhole are still present, even if sometimes being slightly shifted. This is very evident in more dilute distributions (thicker shells) with lighter masses, but the presence of these modes due to the wormhole can also be noted in denser distributions (narrower shells) with heavier masses surrounding the wormhole. In the latter case however, when considering normal matter, the number of trapped modes (and consequently resonant peaks) is larger, since the new quasibound states within the thick shell introduce new resonant peaks in the transmission probability. An interesting aspect to point out is that, the effects of normal matter in the environment are stronger in the low-frequency regime, while in the high-frequency regime the transmission coefficients are almost the same. In particular, by considering dense shells of matter, the new resonant peaks are mainly present at lower frequencies.
Remarkably, for the phantom matter case, the differences on the transmission coefficients are very subtle, even for heavier values of the shell mass in more dilute configurations. However, in denser distributions of phantom matter, due to the appearance of a narrower peak in the effective potential, higher than the ones inherited from the thin-shell grafting, non-negligible deviations of the transmission probability are noticed.

\subsubsection{Absorption spectra}
\begin{figure*}
\includegraphics[width=\columnwidth]{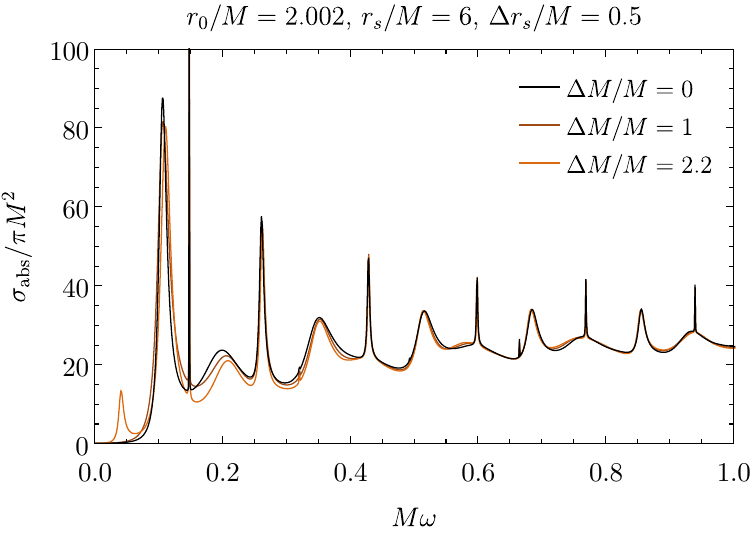}
\includegraphics[width=\columnwidth]{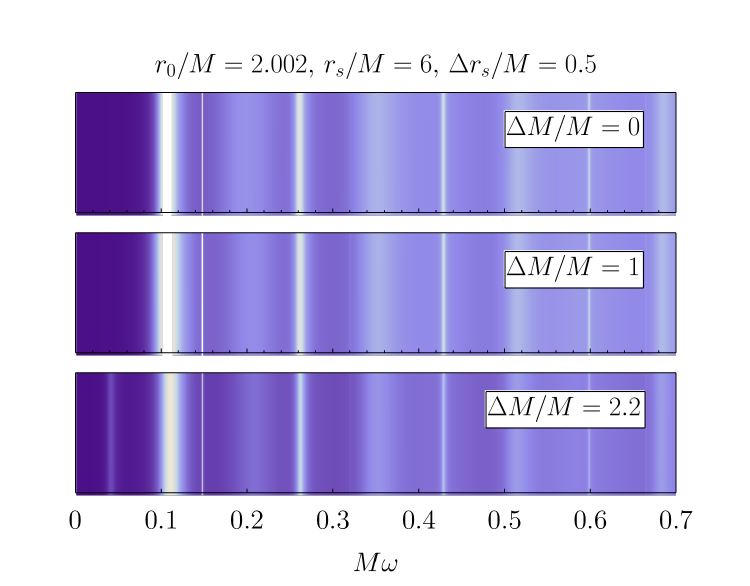}\\
\includegraphics[width=\columnwidth]{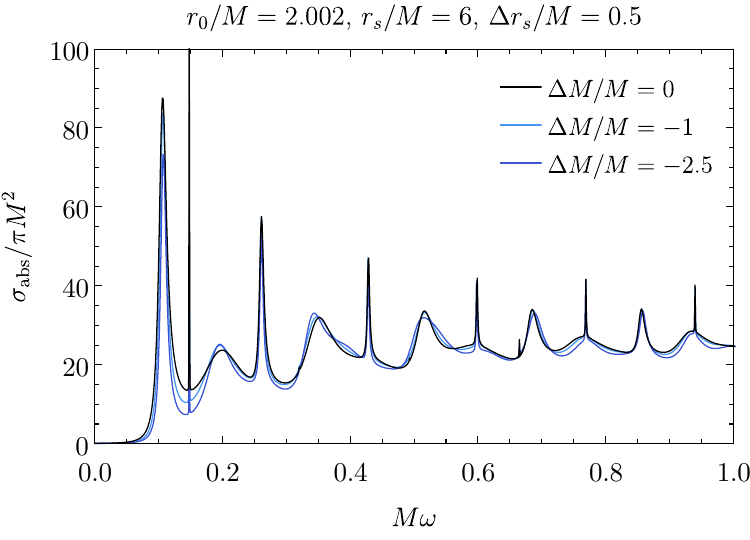}
\includegraphics[width=\columnwidth]{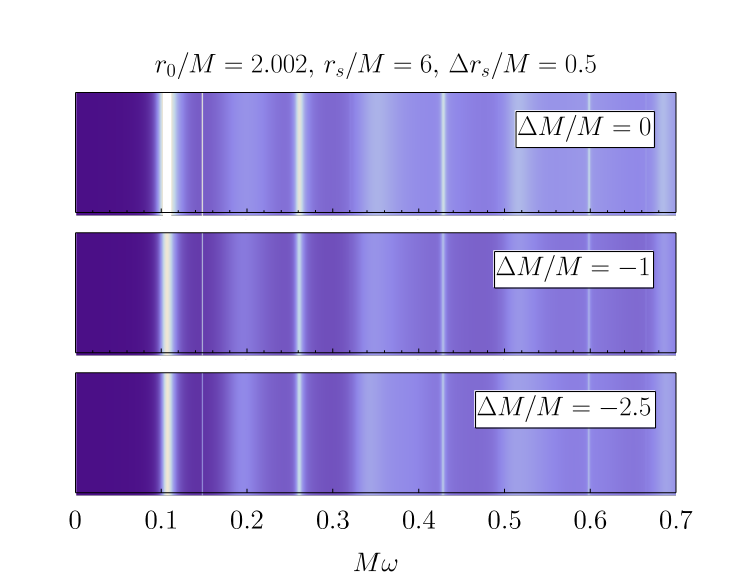}
\caption{Scalar absorption spectra of dirty wormholes. In the left column we exhibit the total absorption cross section for different thick-shell masses and in the right column we show the corresponding absorption bands.}
\label{fig:abs_1}
\end{figure*}

With the transmission probabilities, one can compute the total absorption cross section, $\sigma_{\text{abs}}$, via Eq.~\eqref{eq: Sabs}. In Fig.~\ref{fig:abs_1} we show a selection of our numerical results for the total absorption cross section of dirty wormholes for different thick-shell masses as well as the corresponding absorption bands. In all the configurations considered in Fig.~\ref{fig:abs_1}, the throat of the wormhole has radius $r_0 = 2.002M$, therefore a stable photon sphere is present at the throat and it is surrounded by one unstable photon sphere in each side of the wormhole. As discussed above, the presence of trapped modes around these wormholes is associated to the shape of the effective potential, what introduces resonant peaks in the transmission coefficients (cf. Fig.~\ref{fig:tra_coef_thickness}). These peaks in the transmission probabilities, at specific frequencies, result in the appearance of narrow peaks in the absorption spectra, which are absent in the black hole scenario without surrounding matter.  

The results in Fig.~\ref{fig:abs_1} show that lighter matter distributions in the environment of wormholes have low impact on the total absorption cross section. Deviations are expected when heavier distributions of normal matter are considered, since more trapped modes are present in the spacetime. In these scenarios, new resonant peaks arise changing the number of spectral lines in the absorption cross section. For phantom matter distributions, one expects notable deviations in the total absorption cross section only for extremely heavy and dense configurations, therefore less astrophysically relevant scenarios. The absorption bands shown in Fig.~\ref{fig:abs_1} illustrate how the absorption process is affected by distributions of matter in the environment of wormholes. In particular, one can notice the appearance of new spectral lines when heavier distributions of normal matter are considered.

\begin{figure*}
\includegraphics[width=\columnwidth]{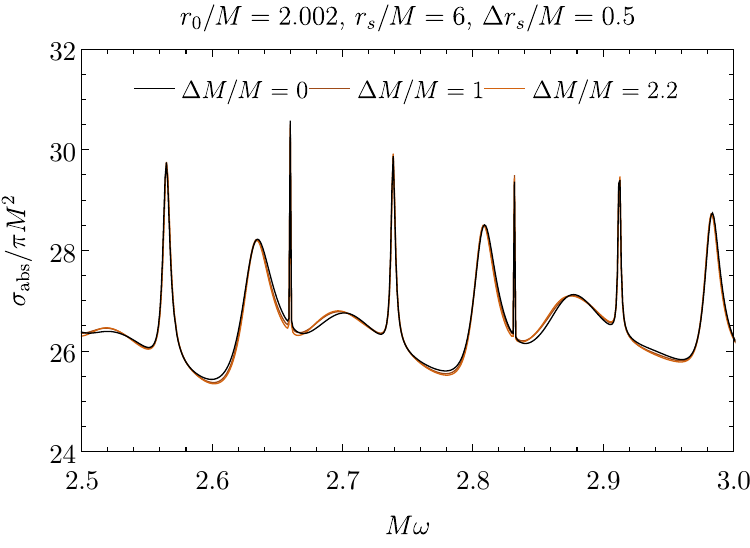}
\includegraphics[width=\columnwidth]{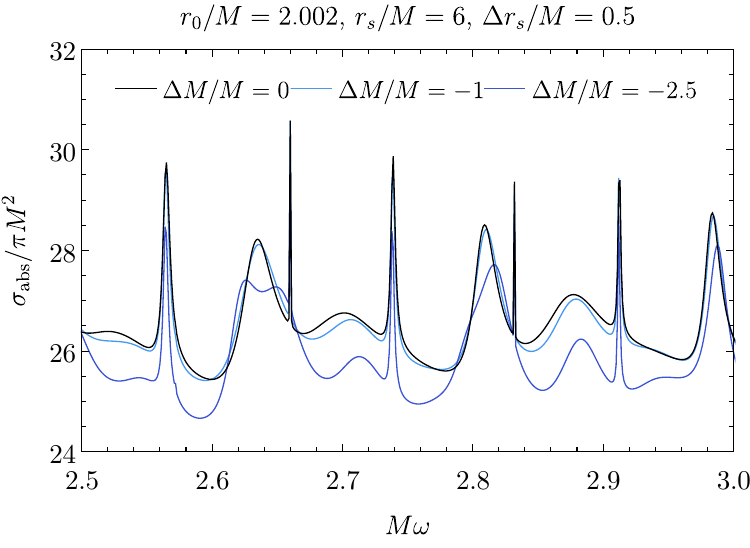}
\caption{High-frequency regime of the total absorption cross section of dirty wormholes. We show the total absorption cross section of some dirty wormholes surrounded by normal matter (right panel), as well as by phantom matter (left panel).}
\label{fig:abs_2}
\end{figure*}

As previously pointed out, we obtained that in the presence of heavier matter distributions, the high-frequency regime of the transmission coefficients of dirty wormholes behave differently, depending on the kind of matter surrounding the wormhole. Our results show that, when surrounded by normal matter, the transmission coefficients do not change much at the high-frequency regime, compared to the case without surrounding matter. When phantom matter is considered, heavier distributions lead to significant changes in the transmission coefficients. Hence, the total absorption cross section of these two scenarios should present notable differences. In Fig.~\ref{fig:abs_2} we show the total absorption cross section of dirty wormholes at moderate-to-high frequencies, where one notices that at high frequencies the total absorption cross section of dirty wormholes surrounded by normal matter is almost the same as thin-shell Schwarzschild wormholes without surrounding matter. On the other hand, depending on their mass, the total absorption cross section with phantom matter is smaller than that of wormholes with or without normal matter. This can be understood from the analysis of the null geodesics effective potential. As previously discussed, when heavier distributions of phantom matter are present, the dominant light ring may change \cite{junior2021can}, changing the shadow area and consequently the eikonal limit of the scalar absorption. Remarkably, even though heavy distributions of surrounding matter may change the absorption cross section at the high-frequency regime, one notices that the location of the spectral lines is preserved. This indicates that, although compact objects may be surrounded by heavy distributions of matter, fingerprints of the central object are still visible in spectral analyses.

\section{Final remarks} \label{sec: FR}
We have investigated the environmental effects on the absorption process of dirty wormholes surrounded by a thick shell of matter. In particular, we studied how the environment can influence the absorption of null particles and massless scalar fields.
The wormhole was constructed via the matching of two Schwarzschild spacetimes. We considered configurations of both normal and phantom matter, consisting of light and dilute as well as heavy and dense distributions surrounding the wormhole.

The presence of the thick shell surrounding the wormhole may sharply modify the null geodesic structure. Depending on the mass of the distribution, both normal and phantom matter may introduce new light rings in the structure of the spacetime. Remarkably, when these new light rings appear they do it in pairs, one of them being stable~\cite{xavier2024traversable}. This occurs due to the conservation of the topological charge associated to the number of light rings in the wormhole spacetime, even when heavier and denser distributions of matter are considered. Moreover, in phantom matter distributions, when the shell is heavier enough, the shadow cast by the dirty wormhole is smaller than that of wormholes with the same mass in empty space. This also implies in a smaller high-frequency absorption cross sections of these configurations.

The existence of new stable light rings around the thick shell of matter indicates that new quasibound states, besides the ones trapped near the throat, can arise. Our investigation of scalar waves propagating in the vicinity of these dirty wormholes have shown that these new quasibound modes may indeed exist. Hence, new resonant peaks can appear in the absorption bands. When present, the new spectral lines are more concentrated at low frequencies. Surprisingly, at high frequencies even though heavy and dense distributions of matter are considered surrounding the wormhole, the position of the spectral lines in the absorption bands are preserved. Therefore, some fingerprints of wormholes are unaffected by their environment, and even when surrounded by heavy shells of matter, one could determine the main characteristic of the central object.
	
\begin{acknowledgments}
The authors would like to acknowledge 
Funda\c{c}\~ao Amaz\^onia de Amparo a Estudos e Pesquisas (FAPESPA), 
Conselho Nacional de Desenvolvimento Cient\'ifico e Tecnol\'ogico (CNPq)
 and Coordena\c{c}\~ao de Aperfei\c{c}oamento de Pessoal de N\'ivel Superior (CAPES) -- Finance Code 001, from Brazil, for partial financial support.  H.~L.~J. thanks the Federal University of Pará for the kind hospitality during the completion of this work.
This research has further been supported by the European Union's Horizon 2020 research and innovation (RISE) programme H2020-MSCA-RISE-2017 Grant No. FunFiCO-777740 and by the European Horizon Europe staff exchange (SE) programme HORIZON-MSCA-2021-SE-01 Grant No. NewFunFiCO-101086251.
\end{acknowledgments}

\bibliography{mybib.bib}
\bibliographystyle{report}

\end{document}